\documentclass[a4paper,fleqn,usenatbib,useAMS]{mnras}
\usepackage{newtxtext,newtxmath}
\usepackage[T1]{fontenc}
\usepackage{ae,aecompl}
\usepackage{epsf}
\usepackage{psfig}
\usepackage{graphicx}
\usepackage{lscape}
\usepackage{natbib}
\usepackage{enumerate}
\usepackage[flushleft]{threeparttable}
\usepackage{soul}
\usepackage{amsmath}
\usepackage{verbatim}
\usepackage{longtable}

\newcommand{\temp}{$T_{eff}$}
\newcommand{\logg}{$log(g)$}
\newcommand{\vt}{$v_t$}
\newcommand{\feh}{$[Fe/H]$}
\newcommand{\chrom}{$R^{\prime}_{HK}$}
\newcommand{\HKindex}{$S_{HK}$}

\newcommand{\nodata}{...}

\title[Stellar Parameters of SEEDS Survey]{The Fundamental Stellar Parameters of FGK Stars in the SEEDS Survey}

\author[Rich et al.]{Evan A. Rich$^{1}$
John P. Wisniewski$^{1}$
Michael W. McElwain$^{2}$
Jun Hashimoto$^{3,4}$
\newauthor
Tomoyuki Kudo$^{5}$,
Nobuhiko Kusakabe$^{4}$,
Yoshiko K. Okamoto$^{6}$,
Lyu Abe$^{7}$,
\newauthor
Eiji Akiyama $^{5}$,
Wolfgang Brandner$^{8}$,
Timothy D. Brandt$^{9}$,
Phillip Cargile$^{10}$,
\newauthor
Joseph C. Carson$^{11}$,
Thayne M Currie$^{5}$,
Sebastian Egner$^{5}$,
Markus Feldt$^{8}$,
\newauthor
Misato Fukagawa$^{12}$,
Miwa Goto$^{2}$,
Carol A. Grady$^{13,14,15}$,
Olivier Guyon$^{5}$,
\newauthor
Yutaka Hayano$^{5}$,
Masahiko Hayashi$^{4}$,
Saeko S. Hayashi$^{5}$,
Leslie Hebb$^{16}$,
\newauthor
Krzysztof G. He{\l}miniak$^{17}$,
Thomas Henning$^{8}$,
Klaus W. Hodapp$^{18}$,
Miki Ishii$^{4}$,
\newauthor
Masanori Iye$^{4}$,
Markus Janson$^{19}$,
Ryo Kandori$^{4}$,
Gillian R. Knapp$^{20}$,
\newauthor
Masayuki Kuzuhara$^{21}$,
Jungmi Kwon$^{22}$,
Taro Matsuo$^{23}$,
Satoshi Mayama$^{24}$,
\newauthor
Shoken Miyama$^{25}$,
Munetake Momose$^{26}$,
Jun-Ichi Morino$^{4}$,
Amaya Moro-Martin$^{27,28}$,
\newauthor
Takao Nakagawa$^{29}$,
Tetsuo Nishimura$^{5}$,
Daehyeon Oh$^{30}$,
Tae-Soo Pyo$^{5}$,
\newauthor
Joshua Schlieder$^{31,8}$,
Eugene Serabyn$^{32}$,
Michael L. Sitko$^{33,34}$,
Takuya Suenaga$^{4,35}$,
\newauthor
Hiroshi Suto$^{4}$,
Ryuji Suzuki$^{4}$,
Yasuhiro H. Takahashi$^{4,22}$,
Michihiro Takami$^{36}$,
\newauthor
Naruhisa Takato$^{5}$,
Hiroshi Terada$^{5}$,
Christian Thalmann$^{37}$,
Daigo Tomono$^{5}$,
\newauthor
Edwin L. Turner$^{9}$,
Makoto Watanabe$^{38}$,
Toru Yamada$^{39}$,
Hideki Takami$^{4}$,
\newauthor
Tomonori Usuda$^{4,24}$,
Motohide Tamura$^{4,22}$
\\ \\
Note: All affiliations are located after the conclusion section of the paper.}

\date{Accepted 2017 August 8. Received 2017 August 7; in original form 2017 February 2}
\pubyear{2017}

\begin{document}
\label{firstpage}
\pagerange{\pageref{firstpage}--\pageref{lastpage}}
\maketitle

\begin{abstract}
Large exoplanet surveys have successfully detected thousands of exoplanets to-date.  Utilizing these detections and non-detections to constrain our understanding of the formation and evolution of planetary systems also requires a detailed understanding of the basic properties of their host stars.  We have determined the basic stellar properties of F, K, and G stars in the Strategic Exploration of Exoplanets and Disks with Subaru (SEEDS) survey from echelle spectra taken at the Apache Point Observatory's 3.5m telescope.  Using ROBOSPECT to extract line equivalent widths and TGVIT to calculate the fundamental parameters, we have computed \temp, \logg, \vt, \feh, chromospheric activity, and the age for our sample.
Our methodology was calibrated against previously published results for a portion of our sample. 
The distribution of \feh\ in our sample is consistent with that typical of the Solar neighborhood. Additionally, we find the ages of most of our sample are $< 500 Myrs$, but note that we cannot determine robust ages from significantly older stars via chromospheric activity age indicators.
The future meta-analysis of the frequency of wide stellar and sub-stellar companions imaged via 
the SEEDS survey will utilize our results to constrain the occurrence of detected co-moving companions with the properties of their host stars.

\end{abstract}


\section{Introduction}

Since the discovery of the first exoplanet surrounding a Sun-like star \citep{mayor1995}, dedicated planet surveys such as, those utilizing the Kepler Space Telescope \citep{borucki2009,borucki2010,borucki2011}, the California Planet Search \citep{howard2010,wright2011} and the Anglo-Australian Telescope planet search \citep{tinney2001,butler2001,wittenmyer2014} have expanded the number of confirmed exoplanets to-date to more than $\sim$3,000 exoplanets (exoplanets.org).  These surveys have yielded sufficient numbers of detections to enable correlations with their host star properties, such as mass and metallicity, to better constrain our understanding of how planets form.  

A variety of studies have sought to identify trends between the frequency of exoplanets and a given host star's fundamental parameters.  Shortly after the 
first detections of exoplanets, it was recognized that there was a trend between the occurrence of Jovian-mass exoplanets and their host star metallicity (eg. \citealt{gonzalez1997, fischer2005}).  More recently, this relation has been extended for Jovian-mass planets surrounding intermediate mass sub-giants to M-dwarf hosts \citet{johnson2010} and to terrestrial-size exoplanets (R$<$1.7 R$_{Earth}$) \citep{wang2015}.  Jovian mass planets are seen to increase in frequency around their host stars from M-dwarf stars to A-dwarfs stars \citep{johnson2010}.
It has also been suggested that the frequency of planets varies inversely with 
the lithium abundance of the host star \citep{israelian2009}, though 
this trend is still hotly debated \citep{carlos2016}.  These trends have been identified for planets detected via radial velocity or transit observations; it remains unclear whether such relationships hold for wide-separation planets detected via direct imaging surveys. 

The majority of exoplanets at small angular separations exhibit correlations with their host stars (e.g. \citealt{johnson2010}), which is expected from the core accretion formation \citep{pollack1996}.  Since it is
unclear whether exoplanets detected at wide separation from their host stars form via core accretion or disc instability \citep{boss2001}, it is 
critical to robustly characterize the fundamental stellar properties of 
large direct imaging surveys to better understand the implications and biases of their detection rates.  Partial characterization of the stellar properties of completed large planet imaging surveys has been performed (e.g. \citealt{nielsen2008} for VLT/NACO and \citealt{biller2013,nielsen2013} for Gemini/NICI surveys); and will likely occur for ongoing surveys using Gemini GPI \citep{macintosh2014} and SPHERE \citep{beuzit2008,vigan2016}.  
The most recent large planet imaging survey to be completed is the Strategic Exploration of Exoplanets and Disks with Subaru (SEEDS) survey \citep{tam09,Tamura2016}, whose primary goal was to survey nearby Solar analogs to search for directly imaged planets and the discs from which they formed.  This survey has announced a number of brown dwarf and exoplanet discoveries, including GJ 504 b \citep{kuzuhara2013}, $\kappa$ And b \citep{carson2013}, GJ 758 B \citep{thalmann2009}, Pleiades HII 3441 b \citep{konishi2016}, and ROXs 42B b \citep{currie2014}.  
Characterizing the fundamental parameters of the host stars of this survey will enable one 
to correlate the observed detections of brown dwarfs and Jovian-mass planets with
the properties of their host stars. 

Fundamental stellar atmospheric parameters such as effective temperature (\temp), surface gravity (\logg), and iron abundance (\feh), can be calculated using a variety of well tested and vetted codes.
For example, MOOG \citep{sneden1973} utilizes plane-parallel atmospheric models to perform 
Local Thermodynamic Equilibrium spectral analysis or synthesis, given a set of equivalent widths (EW) measured from a stellar spectrum and a line list.  
Spectroscopy Made Easy (SME; \citealt{valenti1996,valenti2005,piskunov2017}) uses Kurucz \citep{Castelli2004} or MARCS \citep{gustafsson2008} atmospheric models and line data from the Vienna Atomic Line Database (VALD;  \citealt{Kupka1999,Kupka2000,Ryabchikova1997,Piskunov1995}) to fit synthesized spectra to observed spectra.  
Temperature Gravity microtrubulent Velocity ITerations (TGVIT; \citealt{Takeda2002,Takeda2005}) employs tabulated EWs computed from a grid of atmospheric models with varying atmospheric parameters. In this paper, we have adopted TGVIT to characterize the fundamental properties of the SEEDS survey target list.

We present fundamental atmospheric parameters (\temp, \logg, \feh), microturbulent velocity, chromospheric activity, and age determinations of the FGK stars in the SEEDS survey. In section \ref{sec:obs} we present the observations and reduction methods for our echelle spectra. 
Next, we discuss our methodology for measuring line strengths (section \ref{sec:fundamental}) and then using TGVIT (section \ref{sec:parameters}) to calculate the fundamental stellar parameters from these line strengths.  We compare our analysis with a calibration sample in Section \ref{sec:validating}. We also discuss the chromospheric activity ages (section \ref{sec:chrom}) derived from our spectra. We discuss our results in Section \ref{sec:results}.

\section{Observations and Data Reduction}\label{sec:obs}

We observed 110 F,G,K-type stars in the SEEDS master target list with 
the Astrophysical Research Consortium Echelle Spectrograph (ARCES) on the Astrophysical Research Consortium 3.5 meter telescope at the Apache Point Observatory (APO) \citep{wang2003}.  
ARCES provides R $\sim$ 31,500 spectra that cover the wavelength range of 3500 \AA\ to 
10,200 \AA.
These observations were made between 2010 October 2 to 2016 April 13 at a signal to noise (SNR) at 6000 \AA\ ranging from 83 to 483.  
Table \ref{tbl:obs} list the basic properties of our target sample.

These data were reduced using standard techniques in IRAF.\footnote{IRAF is distributed by the National Optical Astronomy Observatory, which is operated by the Association of Universities for Research in Astronomy (AURA) under a cooperative agreement with the National Science Foundation.}
After bias subtraction and flat fielding, the spectral orders were extracted.  
We utilized ThAr lamp exposures taken after each science observation to perform wavelength calibration on these data, and then applied standard heliocentric velocity corrections.  
We determined that the wavelength range 4478 \AA\ - 6968 \AA\ contained a large number of Fe I and Fe II lines at sufficiently high SNR to extract accurate fundamental stellar parameters.  
Thus we next continuum normalized the orders spanning this wavelength range using \textit{continuum} in IRAF \citep{tody1993,tody1986} and a 3rd-4th order spline function.  
The orders containing these continuum normalized data were then merged into a single-order spectrum.  
The systemic velocity for each source (see Table \ref{tbl:atmo_results}) was computed using an IDL-based program that cross correlated a Solar spectrum with each observation.

\section{Analysis}\label{sec:analysis}

The analysis of our observations of the SEEDS target list is aimed at determining the fundamental
stellar parameters for these stars, such as the effective temperature (\temp), surface gravity (\logg), and iron abundance (\feh), as well as the microturbulent velocity correction factor (\vt). We also compute  broad constraints
on the age of stars in this sample, via measurements of chromospheric activity (\chrom).  We utilize TGVIT \citep{Takeda2002, Takeda2005}, which uses observed equivalent widths of Fe I and Fe II lines to determine the fundamental stellar parameters, as detailed in Section \ref{sec:fundamental}.  Our constraints on the ages of these systems is summarized in 
Section \ref{sec:chrom}.

\subsection{Calculating Line Strengths}\label{sec:fundamental}

FGK dwarfs have rich absorption spectra in the optical bandpass; hence, determining
line strengths for a large number of Fe I and Fe II lines in a large sample size is best achieved using some form of automation.  
We used the C-based program ROBOSPECT v2.12 \citep{Waters2013} to 
determine equivalent widths for absorption lines in our sample.  ROBOSPECT used a log boxcar function to identify the local continuum of the normalized spectrum in discrete windows, and calculated the SNR in this region.  ROBOSPECT identifies absorption lines in the spectrum either via a user supplied line list or by searching for n$\sigma$ variations from 
the local continuum.  The program then fits a functional form to those lines to find their EW.

\begin{figure}
\centering
\includegraphics[width=\columnwidth]{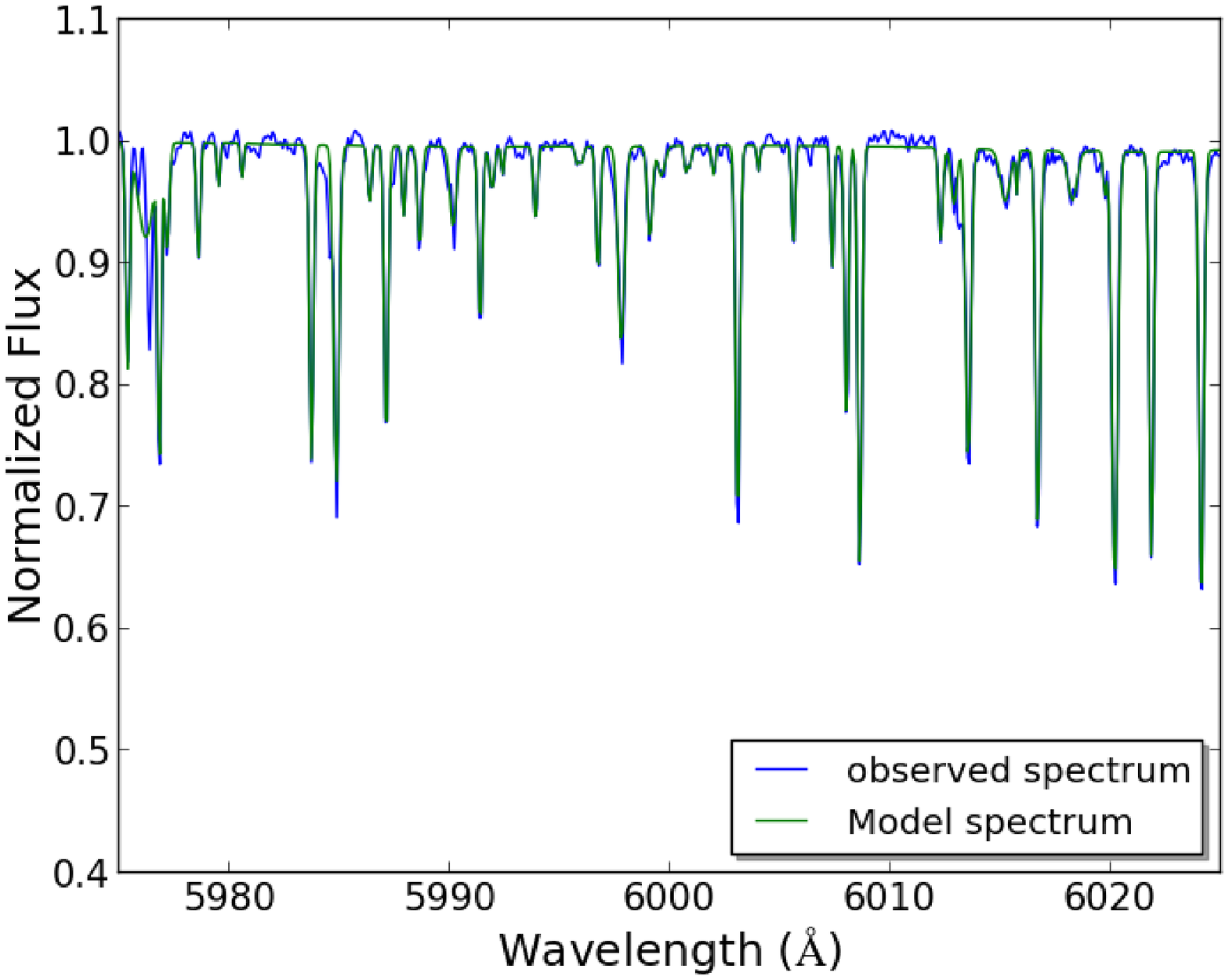}
\caption{This figure plots a sample of HD 172051 spectrum (black) overploted with the ROBOSPECT model spectrum (green).}
\label{fig:spectrum}
\end{figure}

\begin{figure}
\centering
\includegraphics[width=\columnwidth]{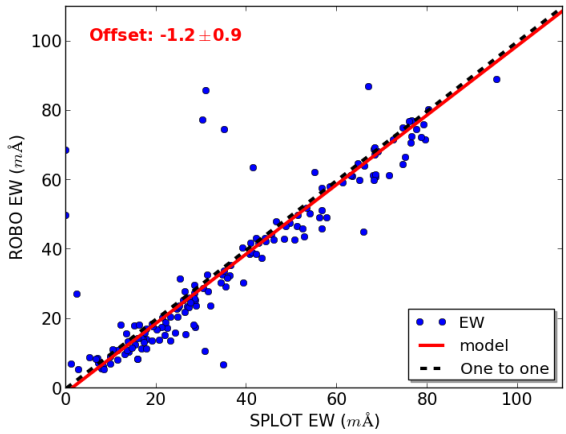}
\caption{This figure plots the EW of HD 172051 measured two different ways. The first with ROBOSPECT (ROBO), the automated line fitting program, and EW measured with SPLOT in IRAF. Note that ROBOSPECT EW values greater than 100 and less than 5 $m$ \AA\ were removed from the sample.}
\label{fig:eqw}
\end{figure}

Through an iterative process, we found that we could achieve qualitative agreement to the observed spectrum by using a window size of 40 m\AA\ for the local continuum normalization, 
used 3$\sigma$ to identify the lines and a gaussian profile to measure the EW values.
In addition to a visual inspection of the synthetic spectrum produced by ROBOSPECT compared 
to the observed spectrum (Figure \ref{fig:spectrum}), we also compared the ROBOSPECT produced EWs 
versus EWs tabulated by hand through the use of \textit{splot} in IRAF.  As shown in 
Figure \ref{fig:eqw}, the EWs determined in an automated fashion via ROBOSPECT mirror 
those computed by hand.  Note that ROBOSPECT tabulated EWs are available as electronic tables in the online version of this manuscript.  We do find a statistically insignificant offset in the EWs determined 
via these two methods (y intercept offset of -1.2 $\pm$ 0.9 $m$\AA\ in 
Figure \ref{fig:eqw}); however, since this offset appears across all of our 
calibration sources it suggests the offset might be systematic and not random 
noise.  As we discuss in Section \ref{sec:atmo_results}, this could lead to an underestimation of \feh.

\subsection{Determining Fundamental Atmospheric Parameters}\label{sec:parameters}

We used the well established FORTRAN based program TGVIT \citep{Takeda2002,Takeda2005} to calculate the fundamental atmospheric parameters (\temp, \logg, \feh) and \vt for our sample.  As described in \citet{Takeda2002}, 
TGVIT utilizes a tabulated grid of model EWs for Fe I and Fe II lines spanning a range of each of the above fundamental 
atmospheric parameters and \vt.  
The code uses a downhill simplex methodology with the tabulated model EWs to iterate to a final set of fundamental stellar parameters for a spectrum. TGVIT is thus different than SME and MOOG based approaches, which calculate the fundamental atmospheric parameters for every combination of line strength in a given spectrum.
TGVIT adopts three criteria 
that are motivated by the effects of the excitation equilibrium, ionization equilibrium, and microturbulence 
on Fe I and Fe II EWs: 

1 - Fe I abundances should not have dependence on the lower excitation potential.

2 - The abundance derived from Fe I should be equal to the abundance derived from Fe II.

3 - The abundances calculated from individual Fe I and Fe II lines in a given star should not have any dependence on the EW.

As described in detail in \citet{Takeda2002}, these three conditions can be represented by a 
single dispersion equation (Equation \ref{eqn:dispersion}).

\begin{equation}
\label{eqn:dispersion}
D^2 \equiv \sigma^2_1 + \sigma^2_2 + (\left< A_1 \right> - \left< A_2 \right> )^2
\end{equation}

Condition 2 can be satisfied where the mean abundance of Fe I ($\left< A_1 \right>$) must equal the mean abundance of Fe II ($\left< A_2 \right>$) thus $\left< A_1 \right> - \left< A_2 \right> = 0$.
Conditions 1 and 3 can be satisfied in the same way, where the deviation of the mean abundance 
of $\left< A_1 \right>$ ($\sigma_1$) and $\left< A_2 \right>$ ($\sigma_2$) must be minimized.
Finally, we follow \citet{Takeda2005} and restrict our analysis to Fe I and Fe II lines whose EW's are less than 100 $m$\AA.

Our initial implementation of TGVIT suggested that the best solution could be biased by a few Fe I and Fe II lines that 
exhibited anomalously high or low EWs. To mitigate this effect, we implemented a bootstrap method similar to that used by \citet{McCarthy2014}.
Our method created 150 unique sets of EWs, each of which were comprised of 90\% of the original Fe I and Fe II lines measured by ROBOSPECT.
We found that our choice of initial fundamental parameters did not affect the results, thus we reused the same initial 
parameter values (\temp = 5000, \logg = 4.0, \vt = 1.0, \feh = 0.0) for our full sample.

We ran each of the 150 unique sets of EWs through TGIVT. Each TGVIT run computed the best fit parameters, calculated the EW residuals (EW$_{data}$- EW$_{TGVIT}$), and identified 
lines that were $\geq$ $2.5\sigma$ outliers. Next, we removed the $\geq$ $2.5\sigma$ outlier lines from the input line list. We then re-ran TGVIT using the initial parameter values.  We performed this iterative rejection procedure for a total of 5 times per unique set of EWs.  
Typically, between 5-20 lines per unique set were removed via this process. Each unique set of EWs provided a single solution of best fit parameter values.
We used the mean of the 150 unique sets to compute the final solution of fundamental stellar parameter values for each star.

We computed uncertainties in our fundamental stellar parameters in two steps. 
First, we adopted the statistical uncertainty calculations within TGVIT described in \citet{Takeda2002}.  The algorithm took steps away from the converged solution in one parameter at a time until one of the three conditions noted above, and re-expressed in equations \ref{eqn:chi_ind}, \ref{eqn:ew_ind}, and \ref{eqn:abd_eq}, was violated. 

\begin{equation}
\label{eqn:chi_ind}
|(\chi_{max} - \chi_{min})\times b| < \sigma_1
\end{equation}

\begin{equation}
\label{eqn:ew_ind}
|(EW_{max} - EW_{min})\times q| < \sigma_1
\end{equation}

\begin{equation}
\label{eqn:abd_eq}
|\left< A_1 \right> - \left< A_2 \right> | < e_1 + e_2
\end{equation}

In the series of inequalities above, the constants $b$ and $q$ represent the slope of a linear-regression fit of the abundance ($A_1$) versus $\chi$ and $A_1$ versus EW respectively and
the constants $e_1$ and $e_2$ are the probable error of the abundance ($\sigma/\sqrt[]{N}$), where N is the number of lines used to calculate the abundance. 
The minimum and maximum values for EW and $\chi$ are taken from the line list and the best fit parameter solution. 
To compute the uncertainty of a parameter, one of the three parameters (\temp, \logg, \vt) is increased until one of the three above inequalities (Equations \ref{eqn:chi_ind}, \ref{eqn:ew_ind}, \ref{eqn:abd_eq}) is violated. The same parameter is then decreased until it violates one of the three above inequalities (Equations \ref{eqn:chi_ind}, \ref{eqn:ew_ind}, \ref{eqn:abd_eq}). The average of the positive and negative differences of the parameter from the best fit value then defines the uncertainty in that parameter. This process is then repeated for the other two parameters. The final uncertainty in the \feh\ abundance was computed by adding the uncertainties in abundance derived from using the accepted range of each of the \temp, \logg, \vt\ parameters in quadrature.  Note that this methodology tested the convergence of isolated parameters.  While \citet{McCarthy2014} demonstrated that the coupled uncertainties between the atmospheric parameters were negligible their analysis was done using spectra of much higher resolution and for only one solar metalicity star. Thus we suggest that the errors we determine should be conservatively viewed as lower limits.

Our use of the bootstrap method allows us to probe how the choice of Fe I and Fe II lines 
influences the converged solutions.  We calculated the standard deviation of each 
parameter over the 150 iterations.  We then added the bootstrap-derived uncertainties to the internally computed TGVIT uncertainties in quadrature.  We note that this final error estimation 
does not take into account any systematical errors.

\subsection{Validating with Calibration Stars} \label{sec:validating}

The fundamental atmospheric properties of stars derived by TGVIT and its precursor program \citep{Takeda2002,Takeda2005} have been robustly compared against a wide variety of techniques 
to compute atmospheric parameters.  As detailed in \citet{Takeda2005}, 
TGVIT has been shown to yield similar parameters as those computed from theoretical 
evolutionary tracks, calculated from B-V \citep{allende1999}, $uvby$ \citep{alonso1996,Olsen1984}, and IR Photometry \citep{ribas2003}, calculated from the wings of H$\beta$ and Mg I b \citep{Fuhrmann1998}, and other spectroscopic analysis programs that invoke 
similar iterative solution approaches outlined in \citet{heiter2003} and \citet{santos2004}.
\citet{Takeda2005} also compared TGVIT results to a collection of atmospheric parameters for 
134 stars compiled from a variety of literature sources by \citet{cayrel2001}.  The offsets between TGVIT and these literature compilations were determined to be \temp = -39 $\pm$ 101 $K$, 
\logg = 0.00 $\pm$ 0.19, and \feh = -0.05 $\pm$ 0.08 \citep{Takeda2005}.  More recently, 
\citet{McCarthy2014} found good agreement between the atmospheric parameters they derived from TGVIT to those derived from SME \citep{valenti2005}, for a sample of 12 stars.

We briefly extend the comparison of TGVIT-derived atmospheric parameters with those derived via other approaches to calibrate our total line list selection procedure and test 
our usage of ROBOSPECT+TGVIT against published literature. Specifically, we utilized 8 stars that were not part of the SEEDS survey, but observed with the same resolution, SNR, and instrument as used in our survey (ARCES at APO).
The first method we used to compute fundamental stellar parameters (referred to as \textit{BPG} in \citealt{wisniewski2012}) used the 2002 version of MOOG \citep{sneden1973}, the one-dimensional plane-parallel model atmospheres interpolated from the ODFNEW grid of ATLAS9 \citep{Castelli2004}, and a line list of $\sim$ 150 Fe I and Fe II lines compiled from the Solar Flux Atlas \citep{kuruca1984}, Utrecht spectral line compilation \citep{moore1966}, and the Vienna Atomic Line Database \citep{Kupka2000,Kupka1999,Ryabchikova1997,Piskunov1995}. 
The second method we used to compute these stellar parameters (referred to as \textit{IAC} in \citealt{wisniewski2012}) also used MOOG \citet{sneden1973}, but with an equivalent width line finding program like ROBOSPECT. 
The third method we used to compute stellar parameters utilized SME \citep{valenti1996,valenti2005}, following the methodology described in \citet{petigura2017}.

\begin{figure*}
\centering
\includegraphics[scale=0.42]{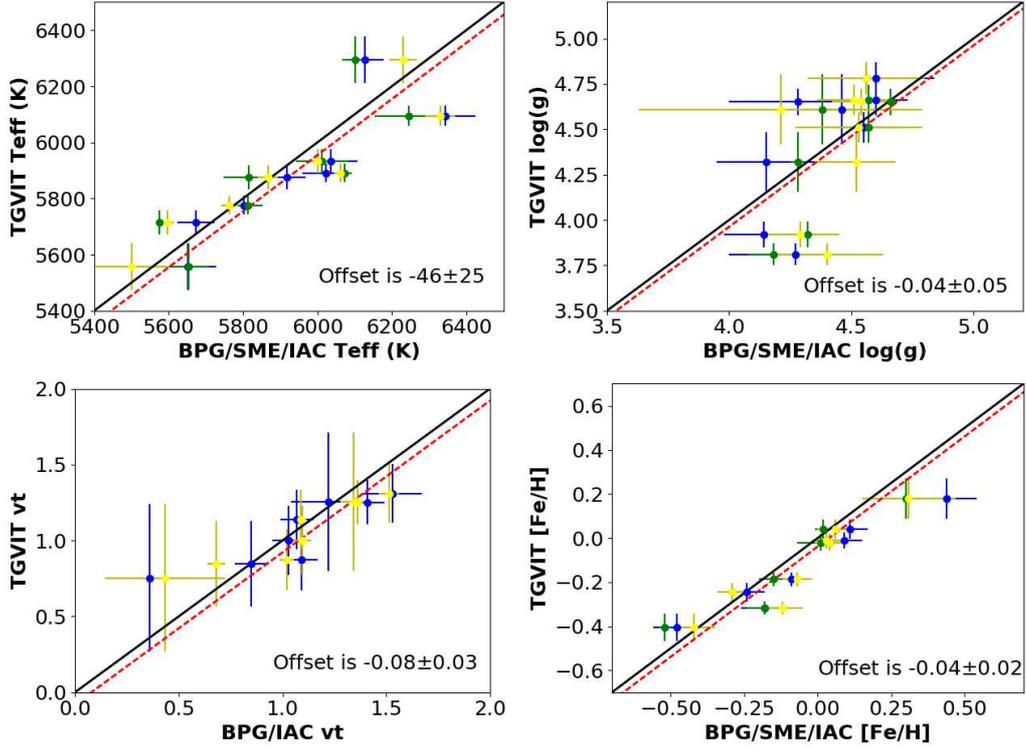}
\caption{Plots the stellar parameters of the 8 calibration stars listed in Table \ref{tbl:calibration} compared to this work's stellar parameters. The three methods utilized are colour coated where SME is blue, ARES is yellow, and BPG is green. Each subplot represents a different stellar parameter plotting the one to one line, and the best-fitting line with each y-axis offset labeled in the subfigure.}
\label{fig:calibrate}
\end{figure*}

We processed our observations of these 8 stars in the same manner as our SEEDS target data, measuring line strengths via ROBOSPECT and computing fundamental parameters via TGVIT, as summarized in Section \ref{sec:parameters}.  
The fundamental parameters derived via our approach and the three methods described above are plotted in Figure \ref{fig:calibrate} and summarized in Table \ref{tbl:calibration}.

To assess the differences between the parameters derived via these three approaches, we fit the data in Figure \ref{fig:calibrate} using the algorithm Orthogonal Distance Regression (ODR) \citep{boggs1990} in \textit{scipy}\footnote{https://www.scipy.org/}, which takes into account uncertainties in both the x and y directions.  The mean differences between TGVIT 
parameters and those derived by the three alternative approaches is $< 2 \sigma$, as seen in 
Figure \ref{fig:calibrate}.  We do note that there is a clustering of \logg \ values, but these still follow a one-to-one relationship within the errors of the parameters.  
These results help demonstrate that our use of ROBOSPECT and TGVIT reproduce the atmospheric parameters derived via MOOG, using the same input dataset, but with different line lists.

We also used these 8 calibration stars to explore the optimal line identification 
procedure to use with ROBOSPECT. 
Optimizing the line identification procedure is important as unidentified lines effect the placement of the continuum, thus influence the final EW output. Note that these additional lines identified outside of the Fe I and Fe II line list from \citet{Takeda2005} are only used internally for continuum placement in ROBOSPECT and are not used for subsequent analysis.
As noted in Section \ref{sec:fundamental}, ROBOSPECT identifies absorption lines in the spectrum either via a user supplied line list or by searching for n$\sigma$ variations from the local continuum.  We found that we were unable to reproduce the atmospheric parameters for our 8 calibration stars derived using other codes (Table \ref{tbl:calibration}) if we provided no line list to ROBOSPECT.  We also noted that using a full line list from VALD \citep{Kupka2000,Kupka1999,Ryabchikova1997,Piskunov1995} required ROSOSPECT to use large computational times.  Thus, we used the Fe I and II line list from \citet{Takeda2005} and allowed ROBOSPECT to automatically additional identify lines by looking for 3 $\sigma$ deviations from the continuum.

\subsection{Chromospheric Activity Ages}\label{sec:chrom}

We computed a measure of chromospheric activity of our sample to help constrain their ages.  We utilized the chromospheric activity-age relationship from \citet{mamajek2008}, shown in equation \ref{eqn:chrom_ages},
where $\tau$ is the age of the star in Gyr and \chrom \, is the chromospheric activity index.  This relationship is based on chromospheric activity levels measured in young stars in clusters, as well as ages for these clusters 
derived from isochronal fitting.  Thus to calculate the ages of our sample stars, we compute the calcium H and K emission line fluxes (3968.47 \AA\ and 3933.66 \AA\ respectively) from our echelle spectra to determine \chrom.

\begin{multline}\label{eqn:chrom_ages}
\log{(\tau)} = -38.053 - 17.912 \log{(R'_{HK})} \\ - 1.6675 \log{(R'_{HK})}^2
\end{multline}
\\

\chrom \, is defined as the luminosity of the Calcium H and K emission lines divided by the total luminosity of the star. We follow \citet{noyes1984} and compute \chrom\ in Equation \ref{eqn:chrom}. Line luminosities, determined by measuring the line emission flux for the H and K lines, are represented by the flux index \HKindex \, (Equation \ref{eqn:chrom_s}), where $N_H$ and $N_K$ are counts from the core of the H and K lines, $N_V$ and $N_R$ are counts from continuum regions, and $\alpha$ and $\beta$ are correction factors. We use literature values of each star's 
$B-V$ magnitude (see Table \ref{tbl:chromospheric}) to represent the continuum contributing to the luminosity of the H and K lines, which is encapsulated in the polynomials $C_{1}$ (Equation \ref{eqn:c1_main}; see \citealt{noyes1984}).  The polynomial $C_2$ in Equation \ref{eqn:c2}, adopted from \citet{noyes1984}, encapsulates the total luminosity of the star.

\begin{equation}
\label{eqn:chrom}
\log_{10}{(R'_{HK})} = \log_{10}{(1.34 \times 10^{-4} S_{HK} 10^{C_{1}} - 10^{C_{2}})}
\end{equation}

\begin{equation}
\label{eqn:chrom_s}
S_{HK} = \alpha \frac{N_H + N_K}{N_V + N_R} + \beta
\end{equation}

\begin{multline}
\label{eqn:c1_main}
C_{1} = 1.13 \times (B-V)^3 - 3.91 \times (B-V)^2 \\ + 2.84 \times (B-V) - 0.47
\end{multline}

\begin{equation}
\label{eqn:c2}
C_2 = -4.898 + 1.918 \times (B-V)^2 - 2.893 \times (B-V)^3
\end{equation}

\begin{figure}
\centering
\includegraphics[width=\columnwidth]{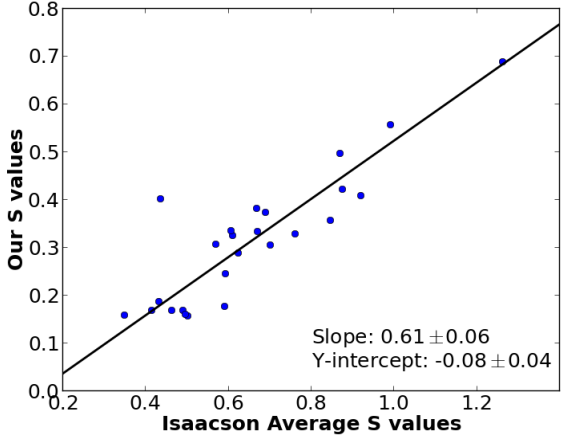}
\caption{\HKindex indices are plotted for \citet{isaacson2010} and our measured \HKindex values. The \HKindex indicies are calculated from measured H and K Ca lines fluxes. The S values from \citet{isaacson2010} are an average of multiple observations of the same object to study jitter, which affects the value of \HKindex that is calculated. We note that our observations have no control for jitter.}
\label{fig:s_values}
\end{figure}

We measured the strength of the calcium H and K emission lines in a 1 \AA\-wide band at 
the line core, following \citet{middelkoop1982}.
We used regions 20 \AA -wide centered at 3891 \AA\ and 4001 \AA\, which are outside of the H and K absorption lines, to measure the local continuum.
$N_V$ and $N_R$ are the average of these continuum locations (3891 \AA\ and 4001 \AA\ respectively).
To compute the normalization ($\alpha$) and the offset ($\beta$) factors, we calibrated our \HKindex\ index (initially with $\alpha$=1 and $\beta$=0) to \HKindex\ index values calculated by \citet{isaacson2010} for 25 stars in common. Figure \ref{fig:s_values} compares our measured S$_{HK}$ index to the average S$_{HK}$ index from \citet{isaacson2010}; the linear relation determined via use 
of the ODR fitting algorithm described in subsection \ref{sec:validating} yielded these normalization and offset factors.  Using the correction factor, we calculated the final $S_{HK}$ index values, the corresponding \chrom \, values, and the resultant ages (see Table \ref{tbl:chromospheric}).  Note that Equation \ref{eqn:chrom_ages} from \citet{mamajek2008} is only valid for $\log_{10}{(R'_{HK})}$ values between -4.0 and -5.1.  
The uncertainties quoted in the chromospheric-activity ages in Table \ref{tbl:chromospheric} are propagated from the uncertainties in the normalization and offset factors ($\alpha$ and $\beta$).

\section{Results}\label{sec:results}

We now derive the fundamental atmospheric parameters, chromospheric activity, and age estimates for our entire sample, outlined in Section \ref{sec:analysis}.
Our results for the fundamental atmospheric parameters are described in Section \ref{sec:atmo_results}. Finally, we estimate the age of our stars by measuring their chromospheric activity in Section \ref{sec:age}.

\subsection{Atmospheric Parameter Results} \label{sec:atmo_results}

We present our atmospheric parameter results using TGVIT in Table \ref{tbl:atmo_results}.  We extracted fundamental parameters for 93 stars that had SNR $\geq 50$, were non-double-lined spectroscopic binaries, and were well within the \temp \, parameter grid of TGVIT.
17 stars could not have their atmospheric parameters robustly determined because they were spectroscopic binaries (3 sources), they could not have their line strengths measured with ROBOSPECT (5 sources), they had too low SNR (4 sources), or TGVIT could not converge on a unique solution (5 sources).  We noticed that ROBOSPECT failed to fit 
the continuum between 4478 to $\sim$ 5500 \AA\ for a subset of early K-type stars. Correspondingly, when data from this spectral range was included in our analysis, the resulting atmospheric parameters derived from TGVIT did not match previously published literature results.  We therefore utilized a spectral window of 5500 - 6968 \AA\ for all early K-type stars and used the full spectral window (4478 - 6968 \AA) for F and G-type stars.  We searched for 
evidence that this reduced spectral bandpass biased the stellar parameters by looking at the dispersion of our results as a function of the number of Fe II lines used, versus those published by \citet{Takeda2002b,Takeda2005}, which also utilizes TGVIT, for the 20 stars common to both surveys, and by \citet{valenti2005}, which uses SME for the 49 stars common to both surveys.
We identified no differences in the dispersion present, above the $3 \sigma$ level, between sources whose parameters were derived from our full spectral windows versus those derived from reduced spectral windows. 

\begin{figure*}
\centering
\includegraphics[scale=0.42]{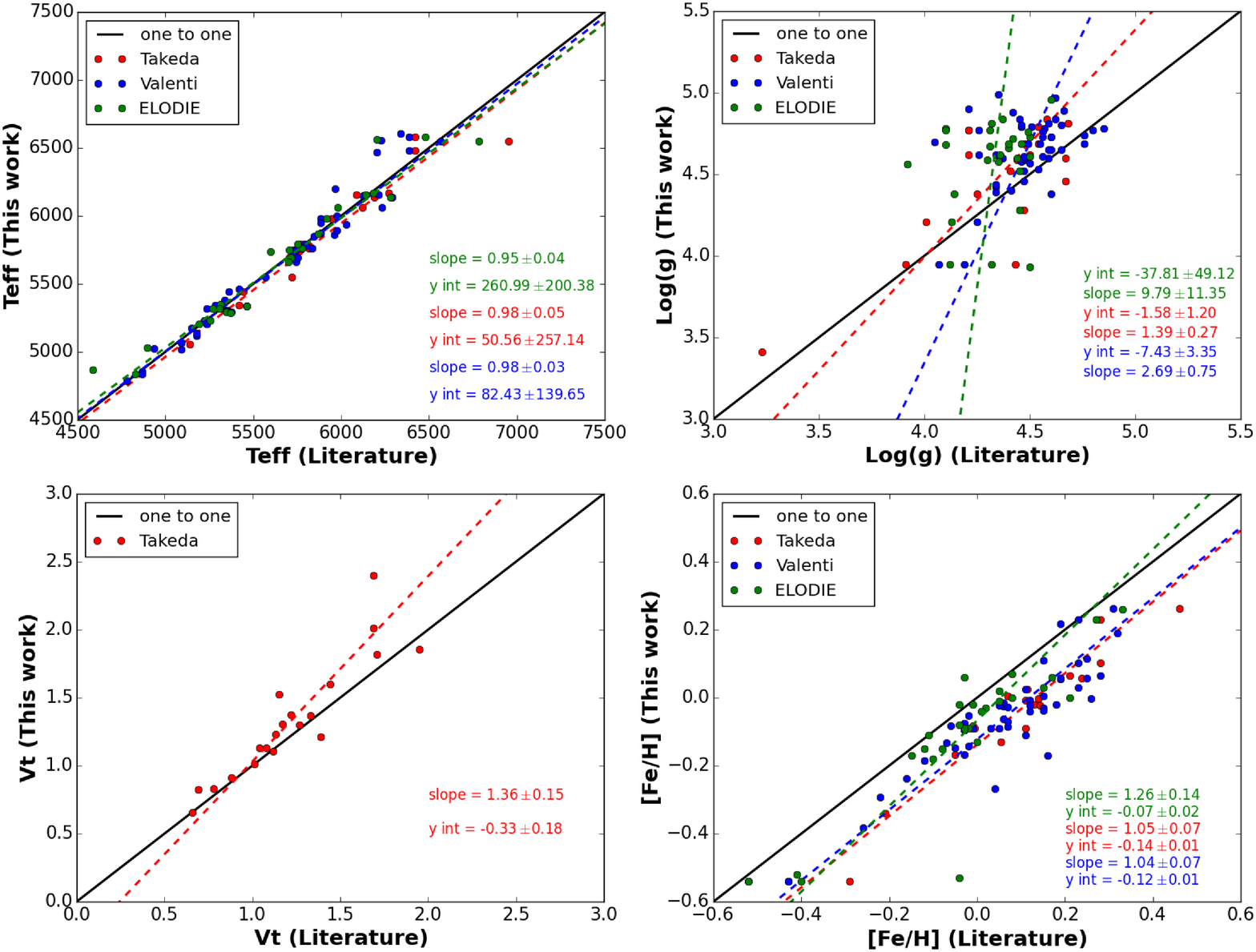}
\caption{The four above panels compare the four fundamental parameters calculated in this work to those calculated in the Takeda et al. (\citeyear{Takeda2005}, \citeyear{Takeda2007}) which has 19 stars in common with our sample (red circles),  \citep{valenti2005} which has 52 stars in common with our sample (blue circles), and ELODIE \citep{prugniel2007} which has 27 stars in common with our sample (green circles). The solid black line represents a one-to-one relation between our values and literature values. The coloured dashed lines represent a linear fit to the corresponding literature values, with the parameters of the best fit shown in coloured text. \temp, \vt, and \feh show linear relations with previous literature values, and \temp and \vt have values consistent with the one-to-one relation. While error bars are not included in this figure due to clarity, Figure \ref{fig:errors} shows the distribution of errors for all four of the atmospheric parameters.}
\label{fig:bias}
\end{figure*}

\begin{figure*}
\centering
\includegraphics[scale=0.42]{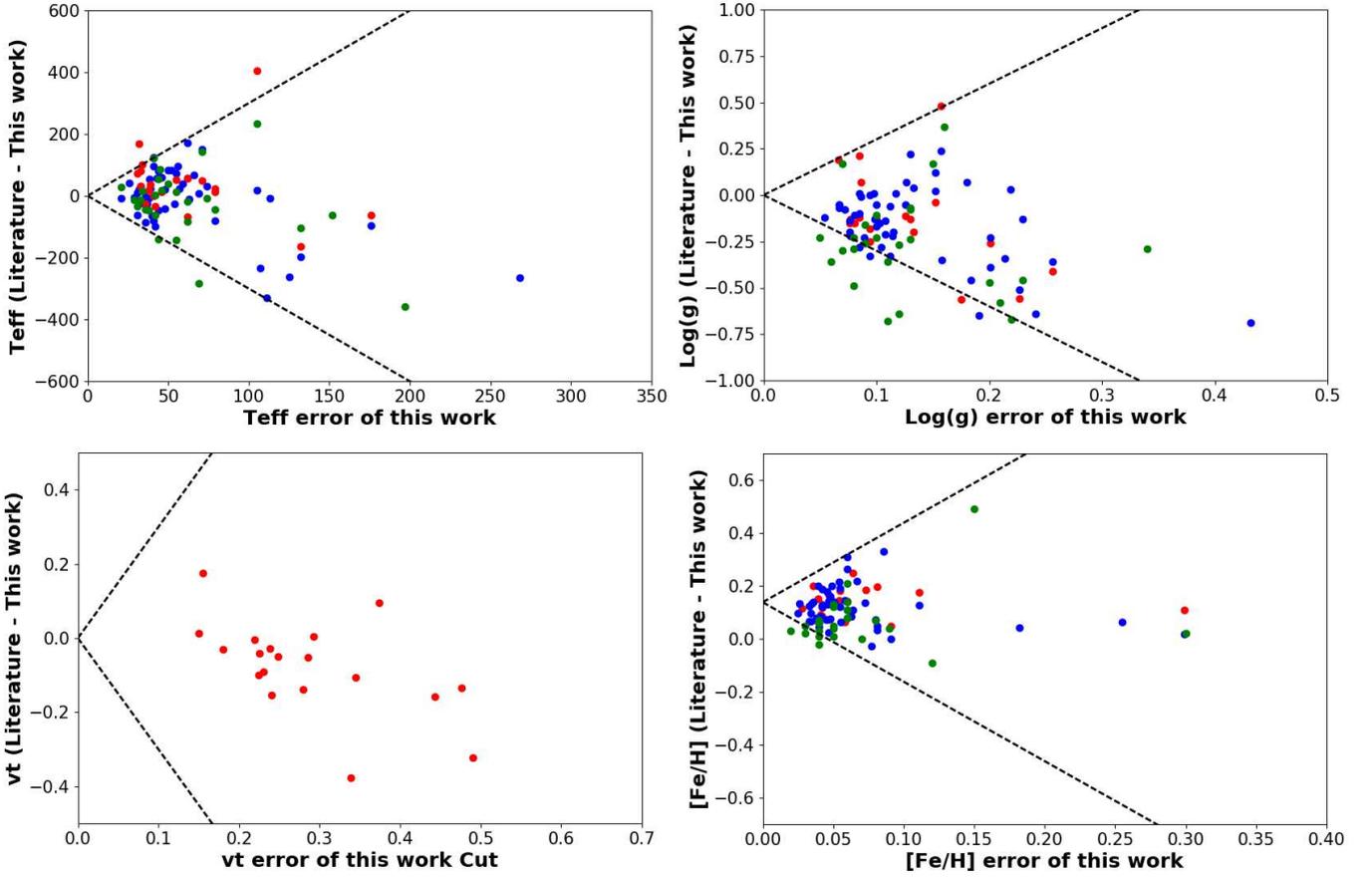}
\caption{Demonstrate the difference between literature values for fundamental stellar 
parameters and those computed via this manuscript are presented as a function of the uncertainties in these parameters. Takeda et al. (\citeyear{Takeda2005}, \citeyear{Takeda2007}) are the red circles, which has 19 stars in common with our sample. \citep{valenti2005} are the blue circles, which has 52 stars in common with our sample. The dashed black lines represent $3 \sigma$, thus anything to the right of the dashed lines is within $3 \sigma$.}
\label{fig:errors}
\end{figure*}

It is common to find systematic offsets in the fundamental atmospheric parameters of stars derived via different methodologies (see e.g. \citealt{Takeda2002b, valenti2005, prugniel2007}).  
We explore the level of these potential offsets in our results, to better enable our results to be utilized by future surveys.  
We compared the amplitude of our fundamental atmospheric parameters to those derived via other spectroscopic methods \citep{Takeda2002b,Takeda2005,valenti2005,prugniel2007}, for an overlapping subset of stars, and fit a linear relation between them using ODR, as shown in Figure \ref{fig:bias}.  
We find that our values of \temp\ are well matched to these previous studies. We find our values of \vt\, are within $3 \sigma$ of Takeda et al. (\citeyear{Takeda2002b,Takeda2005}), which is the only spectroscopic survey of this group that reports the same flavor of \vt\ as our work. Similarly, although there is dispersion between the \logg\ values we derive and those in the literature, these differences are within $3 \sigma$ of the errors (Figure \ref{fig:errors}).

Our derived \feh\ values exhibit minor offsets along the y-axis shown in Figure \ref{fig:bias}. Specifically, our \feh\ values are -0.12 $\pm$0.01 dex smaller than \citet{Takeda2002b,Takeda2005}, -0.14 $\pm$0.01 dex smaller than \citet{valenti2005}, and -0.07 $\pm$ 0.02 smaller than \citet{prugniel2007}.
The slope of the \feh\ offsets is within $2 \sigma$ of unity (see Figure \ref{fig:bias}), indicating that the offsets are a simple constant that could be used to allow one to place our results on the same absolute scale as each of these literature works.
One possible cause of the systematic offset of \feh\ is that ROBOSPECT marginally underestimates EWs as compared to measuring line strengths by-hand, as noted in Section \ref{sec:analysis}.  To further explore this, we increased the EW values of 4 stars by 1 m\AA\ re-ran them through TGVIT, and found an average change in \feh\ of 0.05.  Thus, the marginal underestimation of line strengths by ROBOSPECT can only partially explain these observed offsets.

Finally, to further explore the magnitude and origin of any systematic offsets, we compare the fundamental stellar parameters we computed for our 8 comparison stars (see Figure \ref{fig:calibrate}) to the parameters calculated by \citet{Takeda2002b,Takeda2005}, \citet{valenti2005}, and \citet{prugniel2007} (see Figure \ref{fig:bias}). We find that the slopes and intercepts from the 8 sample stars are within $3 \sigma$ of the slopes and intercepts from our sample of stars in common with \citet{Takeda2002b, Takeda2005} \citet{valenti2005}, and \citet{prugniel2007} for the \temp, \logg, and \vt\ parameters. The \feh\ slopes are also within $3 \sigma$ of one another; however, the y-intercept offsets computed for the sample of stars in common with \citet{Takeda2002b,Takeda2005} and \citet{valenti2005} are $> 3 \sigma$ different compared to that from the 8 comparison stars.

\begin{figure*}
\centering
\includegraphics[scale=0.4]{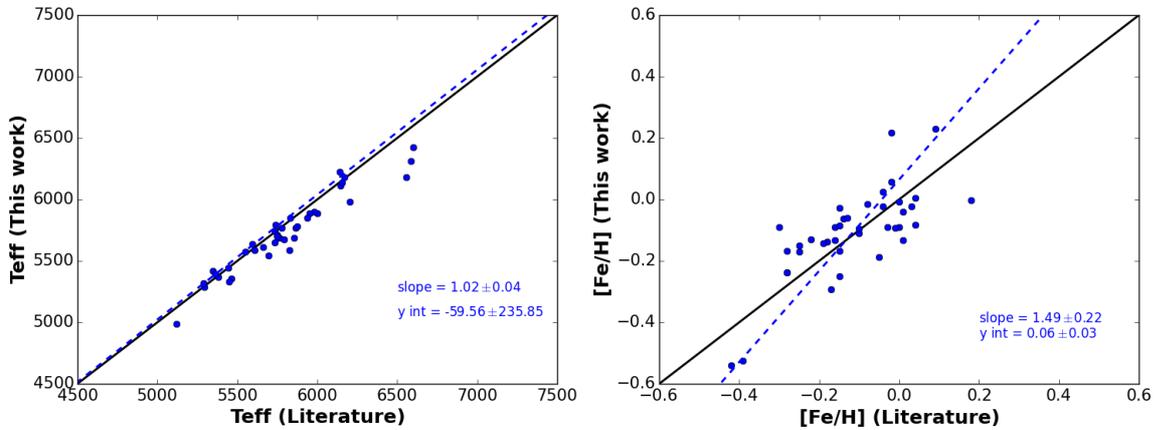}
\caption{Two panels compare the two fundamental parameters calculated in this work to the photometric Copenhagen-Geneva survey \citep{casagrande2011} which has 39 stars in common with our sample. The solid black line represents a one-to-one relation between our values and literature values. The coloured dashed lines represent a linear fit to the corresponding literature values, with the parameters of the best fit shown in coloured text.}
\label{fig:bias_phot}
\end{figure*}

For completeness, we also compared the amplitude of our fundamental atmospheric parameters to those derived via photometric methods in the Geneva-Copenhagen survey of \citet{casagrande2011}, as shown in Figure \ref{fig:bias_phot}.  The offset between our results and those from the Geneva-Copenhagen 
survey is within $3 \sigma$ ($0.06 \pm 0.03$ dex), for common sources.  

Next, we discuss our atmospheric results for GJ 504, a 
G-type star with a directly imaged low-mass companion \citep{kuzuhara2013}.  The age 
of this system, and hence the inferred mass of its wide companion, is a subject of debate in the literature.  \citet{kuzuhara2013} considered a wide range of techniques to assess the age of the system, including gyrochronology, chromospheric activity, x-ray activity, lithium abundances, and isochrones, and adopted a most likely age of $160^{+350}_{-60}$ Myrs.  
\citet{fuhrmann2015} and \citet{dorazi2016} have revisited 
the age estimates for GJ 504, and suggested the system has a much older age, thereby increasing the inferred mass of the wide companion into the brown dwarf regime.  Both \citet{fuhrmann2015} and \citet{dorazi2016} suggest that GJ 504 might have recently engulfed a planetary companion, leading to the unusual rotation and Li abundances observed. Recent atmospheric modeling by \citet{skemer2016} is more consistent with a lower-mass interpretation for the wide companion, hence a younger age estimate for the system, although this work does not exclude the older age hypothesis.   
The fundamental stellar parameters that we compute for GJ 504, 
\temp \, 6063 $\pm$ 62, \logg \, $4.38 \pm 0.13$, and [Fe/H] $0.07 \pm 0.05$ are within the range of stellar parameters for system published by \citet{valenti2005}, \citet{Takeda2007}, \citet{silva2012}, \citet{ramirez2013}, \citet{fuhrmann2015}, and \citet{dorazi2016}. The range of stellar parameters for GJ 504 in some cases exceeds the formal errors quoted for these parameters, likely owing to unrealized calibration offsets between different analysis techniques.  We 
therefore suggest one needs to consider the range of determined fundamental stellar parameters for the system when using these data to determine an age via isochrones. 

\begin{figure*}
\centering
\includegraphics[scale=0.42]{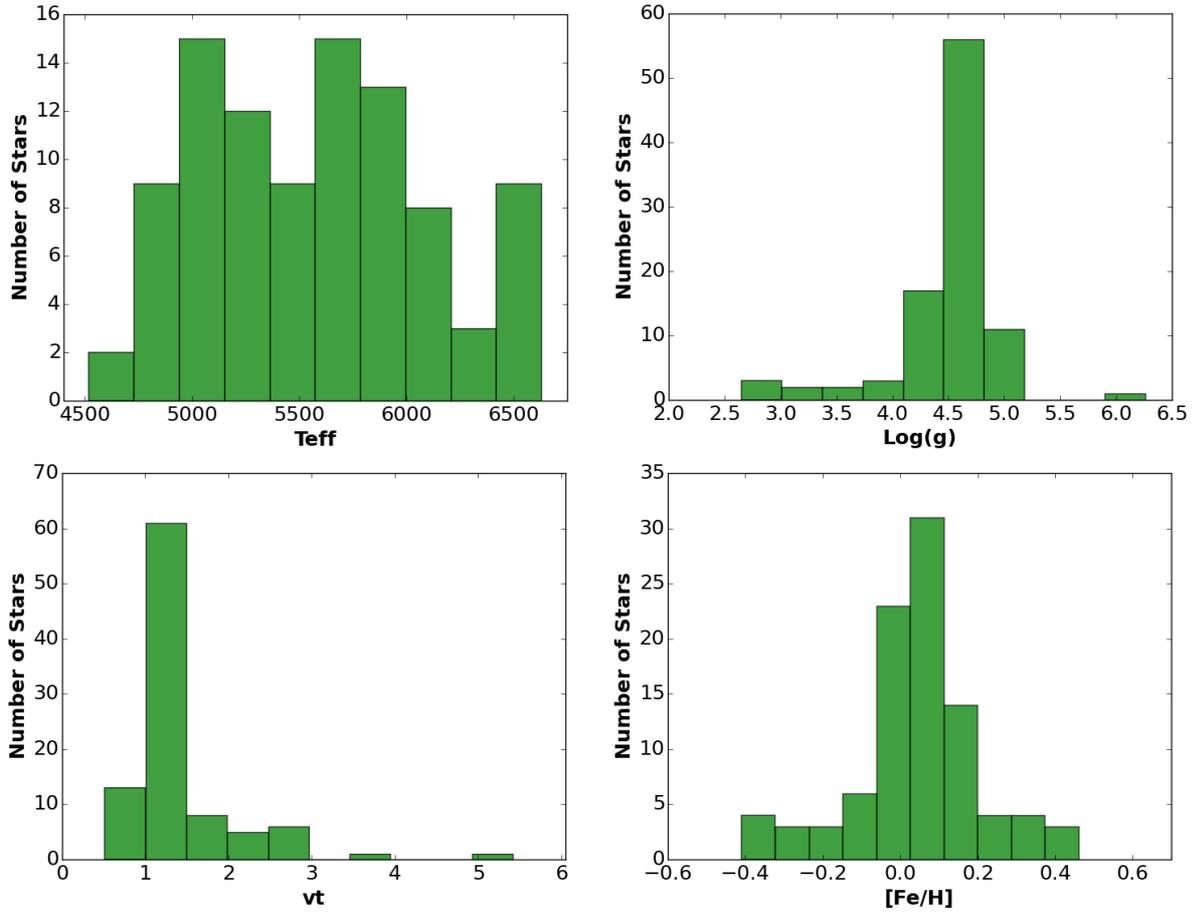}
\caption{Histograms of the four atmospheric parameters determined for stars in our sample using TGVIT (Table \ref{tbl:atmo_results}); \temp\ (upper left), \logg\ (upper right), \vt\ (lower left), and \feh\ (lower right).}
\label{fig:fund_hist}
\end{figure*}

The distribution of our atmospheric parameters for our full sample of stars listed is compiled in Table \ref{tbl:atmo_results}.  Figure \ref{fig:fund_hist} shows histograms of our atmospheric parameters: \temp, \logg, \vt, and \feh. The \temp\ distribution exhibits a fairly uniform distribution across 
FGK space, whereas \vt\ and \logg\ exhibit peaks that are consistent with main sequence stars. Finally, the \feh\ distribution of our sample exhibits a roughly gaussian profile around $0.0 ~dex$, consistent with stars in the solar neighborhood \citep{casagrande2011}. 

\subsection{Chromospheric Activity and Ages}\label{sec:age}

\begin{figure}
\centering
\includegraphics[width=\columnwidth]{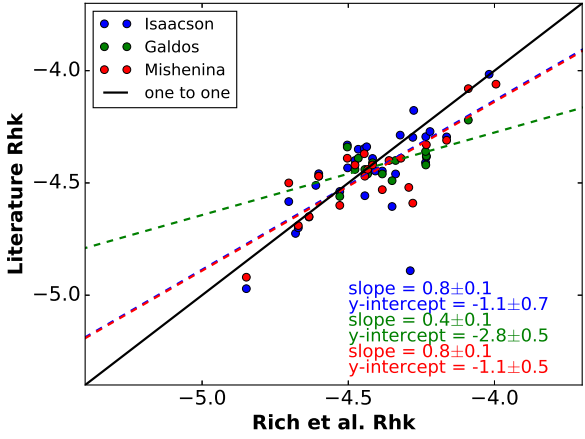}
\caption{This figure compares the previously published literature results of chromospheric activity (\chrom) from \citet{isaacson2010} (blue), \citet{gaidos2000} (green), and \citep{mishenina2012} (red) to our calculated \chrom values (Table \ref{tbl:chromospheric}). The black line is a one-to-one match and the dashed coloured lines are the best linear fit lines using ODR. Results of the fit are printed in coloured text on the bottom right of the figure.}
\label{fig:Rhk}
\end{figure}

We present the chromospheric activity index and associated age estimations for our sample in Table \ref{tbl:chromospheric}. 80 of the 112 stars in our sample had sufficiently strong Ca II H and K emission lines and $B-V$ Values within the validity range of Eq. \ref{eqn:chrom_ages}.

to allow us to calculate \chrom\ values. Our results are consistent with those tabulated by \citet{isaacson2010}, \citet{gaidos2000}, and \citet{mishenina2012} (Figure \ref{fig:Rhk}).
We note that differences from previous published values can result from adopting different $B-V$ values used when calculating \chrom\ and/or intrinsic variability in the level of a star's chromospheric activity (see e.g. \citealt{isaacson2010}).  We determined ages for 51 stars with \chrom\ values within limits of the chromospheric activity-age relation \citep{mamajek2008} as shown in Table \ref{tbl:chromospheric}.
 
\begin{figure}
\centering
\includegraphics[width=\columnwidth]{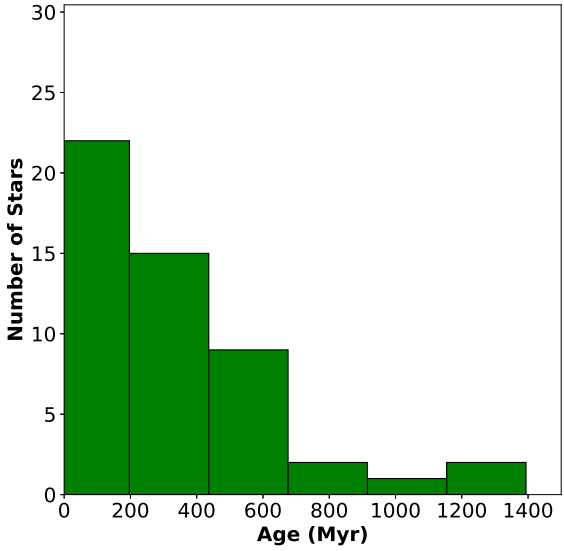}
\caption{Histogram of the calculated ages (Table \ref{tbl:chromospheric}) using the chromospheric activity index.}
\label{fig:chrom_hist}
\end{figure}

Figure \ref{fig:chrom_hist} shows the sample of SEEDS stars for which we were able to derive ages; the majority of the ages are $< 500 Myr$. 
Note that this distribution is not indicative of the complete age distribution of the SEEDS survey, as targets were not selected based on their age. Older stars ($>$ 1.5 Gyr) are outside of the chromospheric activity-age relation, and thus do not have accurate age estimates \citep{mamajek2008}.  

It is important to note that while many stars in the SEEDS sample have ages $< 500 Myr$, their determined ages are still too old to distinguish between core accretion \citep{pollack1996} or disk instability \citep{boss2001} formation scenarios. Planets formed via core accretion are thought to loose most of their heat through the accretion process resulting in "cold-start" planets \citep{spiegel2012}, while planets formed via disk instability retain a lot of their initial heat resulting in "hot-start" planets \citep{spiegel2012}. 
One can distinguish "cold-start" from "hot-start" directly imaged planets via their thermal emission up to an age of $\sim 100 Myr$ old.  While 6 of our stars have ages $<100 Myr$ (see Table \ref{tbl:chromospheric}), the rest of our sample have ages that are either too old or inaccurate to distinguish between cold-start and hot-start formation scenarios for any giant planets they contain \citet{spiegel2012}.

We compare our computed chromospheric ages for stars in known moving groups against the accepted ages for these moving groups. 
\citet{mamajek2008} notes a dispersion for moving group members of 0.25 dex for stars older than 100 Myr and 1 dex for stars less than 100 Myr.  The chromospheric age we determine for the one star in our sample (HD 17925; 42 $\pm$ 12 Myr) in $\beta$ Pic Moving Group is consistent within $2\sigma$ with the estimated cluster age of 23 $\pm$ 3 Myr \citep{mamajek2014}. Similarly, the ages for the three stars in our sample that are part of the Local Association Moving Group, HD 166 (78 $\pm$ 28 Myr), HD 37394 (411 $\pm$ 142 Myr), and HD 206860 (340 $\pm$ 201 Myr), are within $2 \sigma$ of the moving group age of 20-150 Myr \citep{galvez2010}.  The age of our single star located in the Hyades moving group, V401 Hya (205 $\pm$ 95 Myr) is marginally within the 3$\sigma$ range of the moving groups age of 50 $\pm$ 100 Myr \citep{brandt2015}.  The largest dispersion in ages for our stars within moving groups was found in objects located with the Ursa Major moving group.  Although age estimates of this group range from 200-600 Myr (\citealt{eiff2016} and references therein), analysis using MESA models have led to a more recent, precise age of $414 \pm 23$ Myr \citep{jones2015}.  Our ages for HD 43989 (112 $\pm$ 58 Myr), HD 63433 (622 $\pm$ 328 Myr), HD 72985 (79 $\pm$ 36 Myr), and 
HD 135599 (29 $\pm$ 40 Myr) are generally younger than the accepted 
age of the moving group, although including the 0.25 dex dispersion in 
the \citet{mamajek2008} chromospheric age relationship brings all of these age estimates within 3-$\sigma$ agreement except for HD 135599.

Finally, we note that our chromospheric age estimate from our single observation of GJ 504 ($618 \pm 390$ Myr) is consistent with, albeit less precise than, the previously published chromospheric age ($330 \pm 180$ Myr; \citealt{kuzuhara2013}).  We attribute our lower precision to the fact that our age is based on a single epoch of chromospheric activity, whereas the previous chromospheric activity age calculation is based on 30 years of observations.

\section{Conclusion} \label{sec:conclusion}

We have presented the fundamental atmospheric parameters (\temp, \logg, \feh), \vt, chromospheric activity, and age determinations of a subset the FGK stars in the SEEDS survey, based on analysis of high quality, high resolution spectroscopic observations.
We demonstrated the reliability of our methodology by comparing a subset of our results to those published in the literature. 
To aid future comparison of our stellar parameter results with those derived using alternate methodologies, we compile offsets for our computed \feh \ values, (0.06, -0.07, -0.12, -0.14 dex), compared to the respective literature sources (\citealt{casagrande2011}, \citealt{prugniel2007}, Takeda et al. \citeyear{Takeda2002b,Takeda2005}, and \citealt{valenti2005}). Finally, we compared our chromospheric activity and age determinations to previous sources (\citealt{isaacson2010}, \citealt{gaidos2000}, and \citealt{mishenina2012}), and to ages of stars associated with moving groups with known ages.  Our results will aid the interpretation of the frequency of wide stellar and sub-stellar mass companions detected via the SEEDS survey, and comparison of the results of the SEEDS survey with other high-contrast planet and sub-stellar mass imaging surveys.

\section{Acknowledgements}

We thank our referee, Ulrike Heiter, for providing constructive feedback that improved the content and clarity of this manuscript.  This work has made use of the VALD database, operated at Uppsala University, the Institute of Astronomy RAS in Moscow, and the University of Vienna. This research made use of Astropy, a community-developed core Python package for Astronomy \citep{astropy2013}.
This research has made use of the VizieR catalogue access tool, CDS, Strasbourg, France. The original description of the VizieR service was published in A\&AS 143, 23.

$^{1}$Homer L. Dodge Department of Physics, University of Oklahoma, Norman, OK 73071, USA; erich66210@ou.edu, wisniewski@ou.edu \\
$^{2}$Exoplanets and Stellar Astrophysics Laboratory, Code 667, Goddard Space Flight Center, Greenbelt, MD 20771, USA \\
$^{3}$Astrobiology Center of NINS, 2-21-1 Osawa, Mitaka, Tokyo, 181-8588, Japan \\
$^{4}$National Astronomical Observatory of Japan, 2-21-1, Osawa, Mitaka, Tokyo, 181-8588, Japan \\
$^{5}$Subaru Telescope, National Astronomical Observatory of Japan, 650 North A'ohoku Place, Hilo, HI 96720, USA \\
$^{6}$Institute of Astrophysics and Planetary Sciences, Faculty of Science, Ibaraki University, 2-1-1 Bunkyo, Mito, Ibaraki 310-8512, Japan \\
$^{7}$Laboratoire Lagrange (UMR 7293), Universite de Nice-Sophia Antipolis, CNRS, Observatoire de la Cote d'Azur, \\
\ \ 28 avenue Valrose, F-06108 Nice Cedex 2, France \\
$^{8}$Max Planck Institute for Astronomy, K\"{o}nigstuhl 17, D-69117 Heidelberg, Germany \\
$^{9}$Astrophysics Department, Institute for Advanced Study, Princeton, NJ 08540, USA \\
$^{10}$Harvard-Smithsonian Center for Astrophysics, 60 Garden Street, Cambridge, MA 02138, USA
$^{11}$Department of Physics and Astronomy, College of Charleston, 58 Coming St., Charleston, SC 29424, USA \\
$^{12}$Graduate School of Science, Osaka University, 1-1 Machikaneyama, Toyonaka, Osaka 560-0043, Japan \\
$^{13}$Exoplanets and Stellar Astrophysics Laboratory, Code 667, Goddard Space Flight Center, Greenbelt, MD
20771, USA \\
$^{14}$Eureka Scientific, 2452 Delmer, Suite 100, Oakland CA 96002, USA \\
$^{15}$Goddard Center for Astrobiology \\
$^{16}$Department of Physics, Hobart and William Smith Colleges, Geneva, NY 14456, USA \\
$^{17}$Department of Astrophysics, Nicolaus Copernicus Astronomical Center, ul. Rabia\'nska 8, PL-87-100 Toru\'n, Poland \\
$^{18}$Institute for Astronomy, University of Hawaii, 640 N. A'ohoku Place, Hilo, HI 96720, USA \\
$^{19}$Department of Astronomy, Stockholm University, AlbaNova University Center, SE-10691 Stockholm, Sweden \\
$^{20}$Department of Astrophysical Science, Princeton University, Peyton Hall, Ivy Lane, Princeton, NJ 08544, USA \\
$^{21}$Department of Earth and Planetary Sciences, Tokyo Institute of Technology, 2-12-1 Ookayama, Meguro-ku, Tokyo 152-8551, Japan \\
$^{22}$Department of Astronomy, The University of Tokyo, 7-3-1, Hongo, Bunkyo-ku, Tokyo, 113-0033, Japan \\
$^{23}$Department of Astronomy, Kyoto University, Kitashirakawa-Oiwake-cho, Sakyo-ku, Kyoto 606-8502, Japan \\
$^{24}$The Center for the Promotion of Integrated Sciences, The Graduate University for Advanced Studies (SOKENDAI), \\
\ \ Shonan International Village, Hayama-cho, Miura-gun, Kanagawa 240-0193, Japan \\
$^{25}$Hiroshima University, 1-3-2, Kagamiyama, Higashi-Hiroshima 739-8511, Japan \\
$^{26}$College of Science, Ibaraki University, Bunkyo 2-1-1, Mito, 310-8512 Ibaraki, Japan \\
$^{27}$Space Telescope Science Institute, 3700 San Martin Dr., Baltimore, MD 21218, USA \\
$^{28}$Center for Astrophysical Sciences, Johns Hopkins University, Baltimore, MD 21218, USA \\
$^{29}$Department of Space Astronomy and Astrophysics Institute of Space \& Astronautical Science (ISAS) \\
 \ \ Japan Aerospace Exploration Agency (JAXA) 3-1-1 Yoshinodai, Chuo-ku, Sagamihara, Kanagawa 252-5210, Japan \\
$^{30}$National Meteorological Satellite Center, 64-18 Guam-gil, Gawnghyewon-myeon, Jincheon-gun, Chungcheongbuk-do, 27803, Republic of Korea \\
$^{31}$NExScI, California Institute of Technology, Pasadena, CA, 91109, USA \\
$^{32}$Jet Propulsion Laboratory, California Institute of Technology, Pasadena, CA, 91109, USA \\
$^{33}$Department of Physics, University of Cincinnati, Cincinnati, OH 45221, USA \\
$^{34}$Space Science Institute, 475 Walnut Street, Suite 205, Boulder, CO 80301, USA \\
$^{35}$Department of Astronomical Science, The Graduate University for Advanced Studies, 2-21-1, Osawa, Mitaka, Tokyo, 181-8588, Japan \\
$^{36}$Institute of Astronomy and Astrophysics, Academia Sinica, P.O. Box 23-141, Taipei 10617, Taiwan \\
$^{37}$Institute for Astronomy, ETH Zurich, Wolfgang-Pauli-Strasse 27, 8093 Zurich, Switzerland \\
$^{38}$Department of Cosmosciences, Hokkaido University, Kita-ku, Sapporo, Hokkaido 060-0810, Japan \\
$^{39}$Astronomical Institute, Tohoku University, Aoba-ku, Sendai, Miyagi 980-8578, Japan \\

\clearpage

\clearpage

\begin{table*}
\small
\centering
\caption{Basic information for our targets. A full observational list of all F,G, and K stars observed within the SEEDs sample. (1) \citet{gaidos2000}, (2) \citet{esa1997}, (3) \citet{brandt2014}, (4) \citet{houk1988}, (5) \citet{stephenson1986}, (6) \citet{torres2006}, (7) \citet{ehrenreich2011}, (8) \citet{faedi2013}, (9) \citet{bergfors2013}.}
\begin{tabular}{lllccccccc}
\hline
{HD} & {HIP} & {Other} & {RA} & {Dec} & {Date Obs.} & {moving} & {SNR at} & {Spectral} \\
 &  & Name &  &  &  & group & 6000 \AA & Classification\\
\hline
166 & 544 & V439 And & 0:06:37 & 29:01:17 & 2011-03-20 & Local Association$^{(1)}$ & 199.5 & K0V$^{(2)}$ \\
984 & 1134 & BD-08 2400 & 0:14:10 & -7:11:57 & 2010-10-17 & \nodata & 250 & F5$^{(2)}$ \\
1835 & 1803 & BE Cet & 0:22:52 & -12:12:34 & 2012-11-28 & \nodata & 292 & G3V$^{(2)}$ \\
4128 & 3419 & $\beta$ Cet & 0:43:35 & -17:59:12 & 2012-10-24 & \nodata & 310 & K0III$^{(2)}$ \\
4277 & 3589 & BD+54 144 & 0:45:51 & 54:58:40 & 2012-12-09 & \nodata & 70 & F8V$^{(2)}$ \\
4747 & 3850 & GJ 36 & 0:49:27 & -23:12:45 & 2010-10-16 & \nodata & 397 & G8/K0V$^{(2)}$ \\
4813 & 3909 & 19 Cet & 0:50:08 & -10:38:40 & 2011-03-17 & \nodata & 230 & F7IV-V$^{(2)}$ \\
5608 & 4552 & HR 275 & 0:58:14 & 33:57:03 & 2012-10-29 & \nodata & 271 & K0$^{(2)}$ \\
7590 & 5944 & V445 And & 1:16:29 & 42:56:22 & 2012-12-30 & \nodata & 251 & G0$^{(2)}$ \\
7661 & 5938 & EW Cet & 1:16:24 & -12:05:49 & 2012-12-27 & \nodata & 360 & K0V$^{(2)}$ \\
8673 & 6702 & LTT 10515 & 1:26:09 & 34:34:47 & 2012-12-05 & \nodata & 330 & F7V$^{(2)}$ \\
8907 & 6878 & BD+41 283 & 1:28:34 & 42:16:04 & 2011-03-17 & \nodata & 232 & F8$^{(2)}$ \\
8941 & 6869 & BD+16 154 & 1:28:24 & 17:04:45 & 2012-12-30 & \nodata & 483 & F8IV-V$^{(2)}$ \\
9826 & 7513 & $\upsilon$ And & 1:36:48 & 41:24:20 & 2012-12-05 & \nodata & 345 & F8V$^{(2)}$ \\
10700 & 8102 & $\tau$ Cet & 1:44:04 & -15:56:15 & 2012-12-27 & \nodata & 350 & G8V$^{(2)}$ \\
10780 & 8362 & V987 Cas & 1:47:45 & 63:51:09 & 2012-10-04 & \nodata & 352 & K0V$^{(2)}$ \\
11636 & 8903 & $\beta$ Ari & 1:54:38 & 20:48:28 & 2012-12-22 & \nodata & 289 & A5V...$^{(2)}$ \\
12039 & 9141 & DK Cet & 1:57:49 & -21:54:05 & 2012-12-30 & Tuc-Hor$^{(3)}$ & 133 & G3/G5V$^{(2)}$ \\
13507 & 10321 & V450 And & 2:12:55 & 40:40:06 & 2012-12-05 & \nodata & 225 & G0$^{(2)}$ \\
13594 & 10403 & HR 647 & 2:14:03 & 47:29:03 & 2012-12-27 & \nodata & 287 & F5V$^{(2)}$ \\
14067 & 10657 & HR 665 & 2:17:10 & 23:46:04 & 2012-12-05 & \nodata & 257 & G9III$^{(2)}$ \\
14082B & 10679 & BD+28 382B & 2:17:25 & 28:44:42 & 2012-12-09 & \nodata & 272 & G2V$^{(2)}$ \\
16160 & 12114 & HR 753 & 2:36:05 & 6:53:13 & 2012-12-30 & \nodata & 296 & K3V$^{(2)}$ \\
16760 & 12638 & BD+37 604 & 2:42:21 & 38:37:07 & 2012-12-09 & \nodata & 255 & G5$^{(2)}$ \\
17250 & 12925 & BD+04 439 & 2:46:15 & 5:35:33 & 2012-11-28 & \nodata & 168 & F8$^{(2)}$ \\
17925 & 13402 & EP Eri & 2:52:32 & -12:46:11 & 2010-11-19 & beta Pic$^{(3)}$ & 281 & K1V$^{(2)}$ \\
18632 & 13976 & BZ Cet & 3:00:03 & 7:44:59 & 2012-12-27 & \nodata & 218 & G5$^{(2)}$ \\
20630 & 15457 & $\kappa$ Cet & 3:19:22 & 3:22:13 & 2012-11-26 & \nodata & 297 & G5Vvar$^{(2)}$ \\
22781 & 17187 & BD+31 630 & 3:40:50 & 31:49:35 & 2012-11-26 & \nodata & 281 & K0$^{(2)}$ \\
24916 & 18512 & BD-01 565 & 3:57:29 & -1:09:34 & 2012-11-26 & \nodata & 260 & K4V$^{(2)}$ \\
25457 & 18859 & BD-00 423 & 4:02:37 & -0:16:08 & 2010-10-02 & \nodata & 377 & F5V$^{(2)}$ \\
25665 & 19422 & BD+69 238 & 4:09:35 & 69:32:29 & 2010-10-17 & \nodata & 247 & G5$^{(2)}$ \\
25998 & 19335 & V582 Per & 4:08:37 & 38:02:23 & 2012-11-26 & \nodata & 270 & F7V$^{(2)}$ \\
29697 & 21818 & V834 Tau & 4:41:19 & 20:54:05 & 2010-11-19 & \nodata & 258 & K3V$^{(2)}$ \\
30495 & 22263 & IX Eri & 4:47:36 & -16:56:04 & 2012-11-28 & \nodata & 294 & G3V$^{(2)}$ \\
31000 & 22776 & V536 Aur & 4:53:56 & 36:45:27 & 2012-10-24 & \nodata & 207 & G5$^{(2)}$ \\
35850 & 25486 & AF Lep & 5:27:05 & -11:54:04 & 2012-12-27 & \nodata & 348 & F7V$^{(2)}$ \\
36869 & \nodata & AH Lep & 5:34:09 & -15:17:03 & 2012-11-26 & \nodata & 393 & G2V$^{(4)}$ \\
37394 & 26779 & V538 Aur & 5:41:20 & 53:28:52 & 2011-03-17 & Local Association$^{(1)}$ & 350 & K1V$^{(2)}$ \\
37484 & 26453 & GC 7011 & 5:37:40 & -28:37:35 & 2010-11-19 & \nodata & 267 & F3V$^{(2)}$ \\
38393 & 27072 & LTT 2364 & 5:44:28 & -22:26:54 & 2010-10-17 & \nodata & 346 & F7V$^{(2)}$ \\
39587 & 27913 & chi01 Ori & 5:54:23 & 20:16:34 & 2010-11-19 & \nodata & 313 & G0V$^{(2)}$ \\
40774 & 28526 & BD+09 1055 & 6:01:17 & 9:04:20 & 2012-10-04 & \nodata & 190 & G5$^{(2)}$ \\
41593 & 28954 & V1386 Ori & 6:06:40 & 15:32:32 & 2012-12-27 & \nodata & 337 & K0$^{(2)}$ \\
43162 & 29568 & GJ 3389 & 6:13:45 & -23:51:43 & 2012-10-24 & \nodata & 229 & G5V$^{(2)}$ \\
43989 & 30030 & V1358 Ori & 6:19:08 & -3:26:20 & 2012-12-09 & Ursa Major$^{(1)}$ & 217 & G0$^{(2)}$ \\
59747 & 36704 & DX Lyn & 7:33:01 & 37:01:47 & 2012-10-24 & \nodata & 200 & G5$^{(2)}$ \\
60737 & 37170 & GC 10209 & 7:38:16 & 47:44:55 & 2012-12-30 & \nodata & 180 & G0$^{(2)}$ \\ 
61606 & 37349 & V869 Mon & 7:39:59 & -3:35:51 & 2012-12-27 & \nodata & 232.3 & K2V$^{(2)}$ \\ 
63433 & 38228 & V377 Gem & 7:49:55 & 27:21:48 & 2012-12-27 & Ursa Major$^{(1)}$ & 294 & G5IV$^{(2)}$ \\ 
68988 & 40687 & BD+61 1038 & 8:18:22 & 61:27:39 & 2010-10-02 & \nodata & 192 & G0$^{(2)}$ \\ 
69830 & 40693 & LHS 245 & 8:18:24 & -12:37:56 & 2012-12-27 & \nodata & 223 & K0V$^{(2)}$ \\ 
72760 & 42074 & BD-00 2024 & 8:34:32 & -0:43:34 & 2012-12-05 & \nodata & 275 & G5$^{(2)}$ \\ 
72905 & 42438 & 3 UMa & 8:39:12 & 65:01:15 & 2012-12-30 & Ursa Major$^{(3)}$ & 217 & G1.5Vb$^{(2)}$ \\ 
73350 & 42333 & V401 Hya & 8:37:50 & -6:48:25 & 2012-12-30 & Hyades$^{(1)}$ & 231 & G0$^{(2)}$ \\ 
75732 & 43587 & LHS 2062 & 8:52:36 & 28:19:51 & 2010-11-19 & \nodata & 160 & G8V$^{(2)}$ \\ 
76151 & 43726 & BD-04 2490 & 8:54:18 & -5:26:04 & 2012-12-30 & \nodata & 219 & G3V$^{(2)}$ \\ 
77825 & 44526 & BD-15 2685 & 9:04:20 & -15:54:51 & 2012-10-24 & \nodata & 110 & K2V$^{(2)}$ \\ 
79555 & 45383 & HEI 350 & 9:14:55 & 4:26:35 & 2012-12-05 & \nodata & 310 & K0$^{(2)}$ \\ 
\hline
\end{tabular}
\label{tbl:obs}
\end{table*}

\begin{table*}
\small
\centering
\contcaption{Basic information for our targets. A full observational list of all F,G, and K stars observed within the SEEDs sample. (1) \citet{gaidos2000}, (2) \citet{esa1997}, (3) \citet{brandt2014}, (4) \citet{houk1988}, (5) \citet{stephenson1986}, (6) \citet{torres2006}, (7) \citet{ehrenreich2011}, (8) \citet{faedi2013}, (9) \citet{bergfors2013}.}
\begin{tabular}{lllccccccc}
\hline
{HD} & {HIP} & {Other} & {RA} & {Dec} & {Date Obs.} & {moving} & {SNR at} & {Spectral} \\
 &  & Name &  &  &  & group & 6000 \AA & Classification\\
\hline
80715 & 45963 & GJ 1124 & 9:22:26 & 40:12:04 & 2012-12-30 & \nodata & 320 & K2V$^{(2)}$ \\ 
81040 & 46076 & BD+20 2314 & 9:23:47 & 20:21:52 & 2012-10-29 & \nodata & 286 & G0$^{(2)}$ \\ 
82106 & 46580 & LHS 2147 & 9:29:55 & 5:39:19 & 2012-10-24 & \nodata & 311 & K3V$^{(2)}$ \\ 
82443 & 46843 & DX Leo & 9:32:44 & 26:59:19 & 2011-02-14 & \nodata & 297 & K0$^{(2)}$ \\ 
82558 & 46816 & LQ Hya & 9:32:26 & -11:11:05 & 2011-04-17 & \nodata & 176 & K0$^{(2)}$ \\ 
87424 & 49366 & V417 Hya & 10:04:38 & -11:43:47 & 2012-12-05 & \nodata & 205 & K0$^{(2)}$ \\ 
89744 & 50786 & BD+41 2076 & 10:22:11 & 41:13:46 & 2010-10-16 & \nodata & 360 & F7V$^{(2)}$ \\ 
90839 & 51459 & GJ 395 & 10:30:38 & 55:58:50 & 2012-12-05 & \nodata & 321 & F8V$^{(2)}$ \\ 
94765 & 53486 & GY Leo & 10:56:31 & 7:23:19 & 2012-12-05 & \nodata & 195 & K0$^{(2)}$ \\ 
95174 & \nodata & LTT 12936 & 10:59:38 & 25:26:16 & 2012-12-09 & \nodata & 160 & K4V$^{(5)}$ \\ 
96064 & 54155 & HH Leo & 11:04:42 & -4:13:16 & 2012-12-09 & \nodata & 230 & G5$^{(2)}$ \\ 
96167 & 54195 & BD-09 3201 & 11:05:15 & -10:17:29 & 2012-12-22 & \nodata & 149 & G5$^{(2)}$ \\ 
97658 & 54906 & BD+26 2184 & 11:14:33 & 25:42:37 & 2010-10-16 & \nodata & 340 & K1V$^{(2)}$ \\ 
98649 & 55409 & LTT 4199 & 11:20:52 & -23:13:02 & 2012-12-05 & \nodata & 147 & G3/G5V$^{(2)}$ \\ 
102956 & 57820 & BD+58 1340 & 11:51:23 & 57:38:27 & 2012-11-28 & \nodata & 170 & K0III$^{(2)}$ \\ 
105631 & 59280 & BD+41 2276 & 12:09:37 & 40:15:07 & 2011-04-17 & \nodata & 243 & K0V$^{(2)}$ \\ 
109272 & 61296 & HR 4779 & 12:33:34 & -12:49:49 & 2012-12-05 & \nodata & 279 & G8III/IV$^{(2)}$ \\ 
112733 & 63317 & BD+39 2586 & 12:58:32 & 38:16:44 & 2012-12-09 & \nodata & 173 & G5V$^{(2)}$ \\ 
115383 & 64792 & GJ 504 & 13:16:47 & 9:25:27 & 2011-04-17 & \nodata & 280 & G0Vs$^{(2)}$ \\ 
115617 & 64924 & 61 Vir & 13:18:24 & -18:18:40 & 2011-03-20 & \nodata & 249 & G5V$^{(2)}$ \\ 
120136 & 67275 & $\tau$ Boo & 13:47:16 & 17:27:25 & 2012-12-22 & \nodata & 455 & F7V$^{(2)}$ \\ 
120352 & 67412 & BD-00 2743 & 13:48:58 & -1:35:35 & 2011-03-20 & \nodata & 337 & K0$^{(2)}$ \\ 
128167 & 71284 & $\sigma$ Boo & 14:34:41 & 29:44:43 & 2012-12-22 & \nodata & 279 & F3Vwvar$^{(2)}$ \\ 
128311 & 71395 & HN Boo & 14:36:01 & 9:44:48 & 2012-12-05 & \nodata & 242 & K0$^{(2)}$ \\ 
129333 & 71631 & Ek Dra & 14:39:00 & 64:17:30 & 2011-03-20 & \nodata & 291 & F8$^{(2)}$ \\ 
134083 & 73996 & 45 Boo & 15:07:18 & 24:52:09 & 2011-03-20 & \nodata & 480 & F5V$^{(2)}$ \\ 
135599 & 74702 & V379 Ser & 15:15:59 & 0:47:47 & 2010-10-02 & Ursa Major$^{(1)}$ & 251 & K0$^{(2)}$ \\ 
145229 & 79165 & BD+11 2925 & 16:09:27 & 11:34:28 & 2012-12-22 & \nodata & 167 & G0$^{(2)}$ \\ 
152555 & 82688 & BD-04 4194 & 16:54:08 & -4:20:25 & 2010-10-17 & \nodata & 334 & G0$^{(2)}$ \\ 
189733 & 98505 & V452 Vul & 20:00:44 & 22:42:39 & 2012-12-30 & \nodata & 208 & G5$^{(2)}$ \\ 
199665 & 103527 & 18 Del & 20:58:26 & 10:50:21 & 2012-12-30 & \nodata & 246 & G6III$^{(2)}$ \\ 
202575 & 105038 & LTT 16242 & 21:16:33 & 9:23:38 & 2011-04-17 & \nodata & 302 & K2$^{(2)}$ \\ 
206466 & 107146 & BD+08 3000 & 12:19:07 & 16:32:54 & 2010-10-17 & \nodata & 228 & K2$^{(2)}$ \\ 
206860 & 107350 & HN Peg & 21:44:31 & 14:46:19 & 2010-10-16 & Local Assocation$^{(1)}$ & 254 & G0V$^{(2)}$ \\ 
210667 & 109527 & V446 Lac & 22:11:12 & 36:15:23 & 2012-12-30 & \nodata & 206 & K0$^{(2)}$ \\ 
212698 & \nodata & 53 Aqr A & 22:26:34 & -16:44:32 & 2012-11-28 & \nodata & 333 & G2V$^{(6)}$ \\ 
213845 & 111449 & NLTT 54210 & 22:34:42 & -20:42:30 & 2012-12-27 & \nodata & 247 & F7V$^{(2)}$ \\ 
217343 & 113579 & GC 32053 & 23:00:19 & -26:09:13 & 2012-12-27 & \nodata & 231 & G3V$^{(2)}$ \\ 
217813 & 113829 & MT Peg & 23:03:05 & 20:55:07 & 2012-11-28 & \nodata & 266 & G5V$^{(2)}$ \\ 
220182 & 115331 & V453 And & 23:21:37 & 44:05:52 & 2012-10-04 & \nodata & 215 & K1V$^{(2)}$ \\ 
222582 & 116906 & BD-06 6262 & 23:41:52 & -5:59:09 & 2012-11-28 & \nodata & 253 & G5$^{(2)}$ \\ 
283750 & 21482 & V833 Tau & 4:36:48 & 27:07:56 & 2010-10-02 & \nodata & 302 & K2$^{(2)}$ \\ 
\nodata & 36357 & V376 Gem & 7:29:02 & 31:59:38 & 2012-10-24 & \nodata & 135 & K2V$^{(2)}$ \\ 
\nodata & 97657 & HatP 11 & 19:50:50 & 48:04:51 & 2012-12-27 & \nodata & 158 & K5V$^{(2)}$ \\ 
\nodata & 115162 & BD+41 4749 & 23:19:40 & 42:15:10 & 2012-12-27 & \nodata & 260 & G0$^{(2)}$ \\ 
\nodata & \nodata & BD+05 4576 & 20:39:55 & 6:20:12 & 2012-12-27 & \nodata & 99 & K7V$^{(5)}$ \\ 
\nodata & \nodata & HatP 13 & 8:39:32 & 47:21:07 & 2012-10-29 & \nodata & 160 & G4$^{(7)}$ \\ 
\nodata & \nodata & HatP 17 & 21:38:09 & 30:29:19 & 2012-12-30 & \nodata & 154 & \nodata \\ 
\nodata & \nodata & HatP 30 & 8:15:48 & 5:50:12 & 2012-11-28 & \nodata & 136 & \nodata \\ 
\nodata & \nodata & HatP 6 & 23:39:06 & 42:27:58 & 2012-12-09 & \nodata & 100 & F8V$^{(8)}$ \\ 
\nodata & \nodata & Wasp 12 & 6:30:33 & 29:40:20 & 2012-11-26 & \nodata & 83 & G0V$^{(9)}$ \\ 
\hline
\end{tabular}
\end{table*}

\begin{table*}
\small
\centering
\caption{Tabulated comparison of our methods to three alternative methods utilizing 8 calibration stars. These results are plotted in Figure \ref{fig:calibrate}. (a) See \citet{petigura2017} for a description of the methods used to calculate these parameters.  (b) See \citet{wisniewski2012} for a description of the methods used to calculate these parameters.  These stars were used to test and calibrate the two stellar characterization pipelines in \citet{wisniewski2012} and \citet{ghezzi2014}, but their computed fundamental stellar parameters were not formally reported in those publications.}
\begin{tabular}{ccccccc}
\hline
{Study Name} & {Name} & {\temp} & {\logg} & {\vt} & {\feh} & {Reference} \\
{} & {} & {(K)} & {log$_{10}$($cm/s^2$)} & {($\frac{km}{sec}$)} & {(dex)} & {} \\
\hline
 & GSC 01240-00945 & 6095.2 $\pm$ 34.8 & 3.81 $\pm$ 0.06 & 1.309 $\pm$ 0.193 & -0.318 $\pm$ 0.030 & \nodata \\
 & HD 20630 & 5776.4 $\pm$ 33.5 & 4.65 $\pm$ 0.073 & 1.137 $\pm$ 0.196 & -0.011 $\pm$ 0.037 & \nodata \\
 & HD 22484 & 5891.1 $\pm$ 32.1 & 3.92 $\pm$ 0.07 & 1.252 $\pm$ 0.147 & -0.187 $\pm$ 0.031 & \nodata \\
This Work & HD 153458 & 5875.9 $\pm$ 42.1 & 4.66 $\pm$ 0.085 & 1.001 $\pm$ 0.226 & 0.042 $\pm$ 0.044 & \nodata \\
 & HD 172051 & 5714.5 $\pm$ 42.9 & 4.78 $\pm$ 0.089 & 0.847 $\pm$ 0.280 & -0.245 $\pm$ 0.042 & \nodata \\
 & HIP 67526 & 5932.9 $\pm$ 42.2 & 4.51 $\pm$ 0.086 & 0.874 $\pm$ 0.201 & -0.020 $\pm$ 0.037 & \nodata \\
 & TYC 1275-27-1 (MC5) & 6294.8 $\pm$ 84.1 & 4.32 $\pm$ 0.164 & 1.255 $\pm$ 0.458 & -0.405 $\pm$ 0.063 & \nodata \\
 & GSC 03546-01452 & 5556.1 $\pm$ 83.1 & 4.61 $\pm$ 0.193 & 0.753 $\pm$ 0.486 & 0.178 $\pm$ 0.091 & \nodata \\
\hline
 & GSC 01240-00945 & 6246 $\pm$ 92 & 4.18 $\pm$ 0.10 & \nodata & -0.18 $\pm$ 0.06 & (a) \\
 & HD 20630 & 5814 $\pm$ 38 & 4.66 $\pm$ 0.03 & \nodata & 0.03 $\pm$ 0.02 & (a) \\
 & HD 22484 & 6074 $\pm$ 18 & 4.32 $\pm$ -0.09 & \nodata & -0.15 $\pm$ 0.05 & (a) \\
SME & HD 153458 & 5816 $\pm$ 68 & 4.57 $\pm$ 0.11 & \nodata & 0.02 $\pm$ 0.03 & (a) \\
 & HD 172051 & 5576 $\pm$ 9 & 4.56 $\pm$ 0.01 & \nodata & -0.29 $\pm$ 0.01 & (a) \\
 & HIP 67526 & 6013 $\pm$ 72 & 4.57 $\pm$ 0.14 & \nodata & 0.01 $\pm$ 0.08 & (a) \\
 & TYC 1275-27-1 (MC5) & 6101 $\pm$ 34 & 4.28 $\pm$ 0.03 & \nodata & -0.52 $\pm$ 0.02 & (a) \\
 & GSC 03546-01452 & 5654 $\pm$ 55 & 4.38 $\pm$ 0.14 & \nodata & 0.30 $\pm$ 0.05 & (a) \\
\hline
 & GSC 01240-00945 & 6330 $\pm$ 40 & 4.40 $\pm$ 0.23 & 1.513 $\pm$ 0.05 & -0.12 $\pm$ 0.07 & \citealt{wright2013} \\
 & HD 20630 & 5764 $\pm$ 22 & 4.54 $\pm$ 0.12 & 1.086 $\pm$ 0.029 & 0.03 $\pm$ 0.05 & (b) \\
 & HD 22484 & 6063 $\pm$ 19 & 4.29 $\pm$ 0.16 & 1.361 $\pm$ 0.024 & -0.07 $\pm$ 0.05 & (b) \\
IAC & HD 153458 & 5867 $\pm$ 27 & 4.51 $\pm$ 0.15 & 1.096 $\pm$ 0.042 & 0.06 $\pm$ 0.06 & (b) \\
 & HD 172051 & 5596 $\pm$ 19 & 4.56 $\pm$ 0.24 & 0.682 $\pm$ 0.043 & -0.29 $\pm$ 0.05 & (b) \\
 & HIP 67526 & 6000 $\pm$ 24 & 4.53 $\pm$ 0.26 & 1.021 $\pm$ 0.035 & 0.04 $\pm$ 0.05 & \citealt{jiang2013} \\
 & TYC 1275-27-1 (MC5) & 6230 $\pm$ 37 & 4.52 $\pm$ 0.16 & 1.343 $\pm$ 0.065 & -0.42 $\pm$ 0.07 & \citealt{ghezzi2014} \\
 & GSC 03546-01452 & 5502 $\pm$ 100 & 4.21 $\pm$ 0.58 & 0.433 $\pm$ 0.290 & 0.31 $\pm$ 0.16 & \citealt{delee2013} \\
\hline
 & GSC 01240-00945 & 6344 $\pm$ 81 & 4.27 $\pm$ 0.27 & 1.53 $\pm$ 0.14 & -0.18 $\pm$ 0.08 & \citealt{wright2013} \\
 & HD 20630 & 5803.0 $\pm$ 43.0 & 4.28 $\pm$ 0.28 & 1.07 $\pm$ 0.08 & 0.09 $\pm$ 0.06 & (b) \\
 & HD 22484 & 6023.0 $\pm$ 63.0 & 4.14 $\pm$ 0.16 & 1.41 $\pm$ 0.08 & -0.09 $\pm$ 0.06 & (b) \\
 BPG & HD 153458 & 5918.0 $\pm$ 50.0 & 4.60 $\pm$ 0.13 & 1.03 $\pm$ 0.08 & 0.11 $\pm$ 0.06 & (b) \\
 & HD 172051 & 5674.0 $\pm$ 50.0 & 4.60 $\pm$ 0.24 & 0.85 $\pm$ 0.08 & -0.24 $\pm$ 0.06 & (b) \\
 & HIP 67526 & 6037 $\pm$ 71 & 4.55 $\pm$ 0.15 & 1.09 $\pm$ 0.08 & 0.04 $\pm$ 0.06 & \citealt{jiang2013} \\
 & TYC 1275-27-1 (MC5) & 6127.0 $\pm$ 50.0 & 4.15 $\pm$ 0.2 & 1.22 $\pm$ 0.18 & -0.48 $\pm$ 0.08 & \citealt{ghezzi2014} \\
 & GSC 03546-1452 & 5652.0 $\pm$ 75 & 4.46 $\pm$ 0.16 & 0.36 $\pm$ 0.20 & 0.44 $\pm$ 0.10 & \citealt{delee2013} \\

\hline
\end{tabular}
\label{tbl:calibration}
\end{table*}

\begin{table*}
\small
\centering
\caption{Chromospheric Activity Index and Ages}
\begin{tabular}{lllccccc}
\hline
{HD} & {HIP} & {Other Name} & {B-V} & {B-V references} & {\HKindex} & {\chrom} & {Age (Myr)} \\
\hline
166 & 544 & V439 And & 0.75 & \citep{zboril1998} & 0.607 & -4.23 $\pm$ 0.04 & 78 $\pm$ 28 \\ 
984 & 1134 & BD-08 2400 & 0.5 & \citep{hog2000} & 0.317 & -4.36 $\pm$ 0.08 & 216 $\pm$ 136 \\ 
1835 & 1803 & BE Cet & 0.66 & \citep{ducati2002} & 0.356 & -4.43 $\pm$ 0.07 & 384 $\pm$ 198 \\ 
4128 & 3419 & $\beta$ Cet & 1.01 & \citep{ducati2002} & 0.345 & -4.79 $\pm$ 0.06 &  \nodata \\ 
4747 & 3850 & GJ 36 & 0.772 & \citep{koen2010} & 0.277 & -4.68 $\pm$ 0.09 &  \nodata \\ 
4813 & 3909 & 19 Cet & 0.5 & \citep{herbig1965} & 0.145 & -4.96 $\pm$ 0.30 &  \nodata \\ 
5608 & 4552 & HR 275 & 0.991 & \citep{jofre2015} & 0.278 & -4.87 $\pm$ 0.07 &  \nodata \\ 
7590 & 5944 & V445 And & 0.58 & \citep{hog2000} & 0.342 & -4.38 $\pm$ 0.08 & 265 $\pm$ 151 \\ 
7661 & 5938 & EW Cet & 0.77 & \citep{hog2000} & 0.566 & -4.29 $\pm$ 0.04 & 124 $\pm$ 45 \\ 
8673 & 6702 & LTT 10515 & 0.47 & \citep{masana2006} & 0.155 & -4.85 $\pm$ 0.15 &  \nodata \\ 
8907 & 6878 & BD+41 283 & 0.49 & \citep{hog2000} & 0.307 & -4.37 $\pm$ 0.08 & 236 $\pm$ 153 \\ 
8941 & 6869 & BD+16 154 & 0.52 & \citep{hog2000} & 0.116 & -5.30 $\pm$ 0.63 &  \nodata \\ 
9826 & 7513 & $\upsilon$ And & 0.54 & \citep{ducati2002} & 0.186 & -4.77 $\pm$ 0.19 &  \nodata \\ 
10700 & 8102 & $\tau$ Cet & 0.72 & \citep{ducati2002} & 0.218 & -4.80 $\pm$ 0.14 &  \nodata \\ 
10780 & 8362 & V987 Cas & 0.81 & \citep{ducati2002} & 0.296 & -4.67 $\pm$ 0.08 &  \nodata \\ 
13507 & 10321 & V450 And & 0.67 & \citep{hog2000} & 0.379 & -4.41 $\pm$ 0.07 & 317 $\pm$ 156 \\ 
14067 & 10657 & HR 665 & 1.04 & \citep{oja1991} & 0.185 & -5.12 $\pm$ 0.11 &  \nodata \\ 
14082B & \nodata & BD+28 382B & 0.59 & \citep{casagrande2011} & 0.436 & -4.25 $\pm$ 0.06 & 93 $\pm$ 46 \\ 
16160 & 12114 & HR 753 & 0.98 & \citep{alonso1996} & 0.298 & -4.83 $\pm$ 0.07 &  \nodata \\ 
16760 & 12638 & BD+37 604 & 0.69 & \citep{hog2000} & 0.276 & -4.62 $\pm$ 0.10 &  \nodata \\ 
17250 & 12925 & BD+04 439 & 0.52 & \citep{hog2000} & 0.326 & -4.36 $\pm$ 0.08 & 220 $\pm$ 135 \\ 
17925 & 13402 & EP Eri & 0.86 & \citep{ducati2002} & 0.898 & -4.16 $\pm$ 0.03 & 42 $\pm$ 12 \\ 
18632 & 13976 & BZ Cet & 0.953 & \citep{koen2010} & 0.750 & -4.36 $\pm$ 0.03 & 222 $\pm$ 56 \\ 
20630 & 15457 & $\kappa$ Cet & 0.67 & \citep{ducati2002} & 0.336 & -4.48 $\pm$ 0.08 & 521 $\pm$ 274 \\ 
25457 & 18859 & BD-00 423 & 0.5 & \citep{chen2000} & 0.245 & -4.52 $\pm$ 0.11 & 692 $\pm$ 517 \\ 
25665 & 19422 & BD+69 238 & 0.973 & \citep{hog2000} & 0.377 & -4.70 $\pm$ 0.06 &  \nodata \\ 
25998 & 19335 & V582 Per & 0.47 & \citep{chen2000} & 0.206 & -4.61 $\pm$ 0.15 & 1219 $\pm$ 1037 \\ 
30495 & 22263 & IX Eri & 0.64 & \citep{casagrande2011} & 0.296 & -4.53 $\pm$ 0.09 & 737 $\pm$ 433 \\ 
31000 & 22776 & V536 Aur & 0.77 & \citep{hog2000} & 0.478 & -4.37 $\pm$ 0.05 & 248 $\pm$ 95 \\ 
35850 & 25486 & AF Lep & 0.503 & \citep{tagliaferri1994} & 0.404 & -4.22 $\pm$ 0.06 & 70 $\pm$ 38 \\ 
36869 & \nodata & AH Lep & 0.717 & \citep{hog2000} & 0.472 & -4.33 $\pm$ 0.05 & 176 $\pm$ 73 \\ 
37394 & 26779 & V538 Aur & 0.84 & \citep{ducati2002} & 0.479 & -4.44 $\pm$ 0.05 & 411 $\pm$ 142 \\ 
38393 & 27072 & LTT 2364 & 0.47 & \citep{ducati2002} & 0.126 & -5.08 $\pm$ 0.42 &  \nodata \\ 
39587 & 27913 & chi01 Ori & 0.6 & \citep{ducati2002} & 0.323 & -4.44 $\pm$ 0.08 & 393 $\pm$ 229 \\ 
41593 & 28954 & V1386 Ori & 0.825 & \citep{koen2010} & 0.709 & -4.24 $\pm$ 0.04 & 79 $\pm$ 25 \\ 
43162 & 29568 & GJ 3389 & 0.673 & \citep{gaidos2002} & 0.431 & -4.34 $\pm$ 0.06 & 187 $\pm$ 85 \\ 
43989 & 30030 & V1358 Ori & 0.57 & \citep{hog2000} & 0.406 & -4.28 $\pm$ 0.06 & 112 $\pm$ 58 \\ 
59747 & 36704 & DX Lyn & 0.88 & \citep{hog2000} & 0.520 & -4.45 $\pm$ 0.04 & 417 $\pm$ 131 \\ 
60737 & 37170 & GC 10209 & 0.633 & \citep{hog2000} & 0.448 & -4.28 $\pm$ 0.06 & 115 $\pm$ 54 \\ 
61606 & 37349 & V869 Mon & 0.967 & \citep{koen2010} & 0.854 & -4.32 $\pm$ 0.03 & 161 $\pm$ 39 \\ 
63433 & 38228 & V377 Gem & 0.68 & \citep{hog2000} & 0.327 & -4.50 $\pm$ 0.08 & 622 $\pm$ 328 \\ 
68988 & 40687 & BD+61 1038 & 0.65 & \citep{hog2000} & 0.128 & -5.29 $\pm$ 0.49 &  \nodata \\ 
72760 & 42074 & BD-00 2024 & 0.825 & \citep{koen2010} & 0.445 & -4.47 $\pm$ 0.05 & 482 $\pm$ 175 \\ 
72905 & 42438 & 3 UMa & 0.62 & \citep{ducati2002} & 0.476 & -4.23 $\pm$ 0.05 & 79 $\pm$ 36 \\ 
73350 & 42333 & V401 Hya & 0.669 & \citep{koen2010} & 0.419 & -4.35 $\pm$ 0.06 & 205 $\pm$ 96 \\ 
75732 & 43587 & LHS 2062 & 0.87 & \citep{ducati2002} & 0.179 & -5.01 $\pm$ 0.14 &  \nodata \\ 
79555 & 45383 & HEI 350 & 1.042 & \citep{koen2010} & 0.733 & -4.49 $\pm$ 0.03 & 577 $\pm$ 126 \\ 
81040 & 46076 & BD+20 2314 & 0.64 & \citep{oja1987} & 0.228 & -4.70 $\pm$ 0.13 &  \nodata \\ 
82443 & 46843 & DXLeo & 0.77 & \citep{kotoneva2006} & 0.849 & -4.09 $\pm$ 0.03 & 20 $\pm$ 6 \\ 
87424 & 49366 & V417 Hya & 0.913 & \citep{koen2010} & 0.497 & -4.50 $\pm$ 0.04 & 623 $\pm$ 186 \\ 
90839 & 51459 & GJ 395 & 0.488 & \citep{heiter2003} & 0.126 & -5.11 $\pm$ 0.44 &  \nodata \\ 
94765 & 53486 & GY Leo & 0.692 & \citep{koen2010} & 0.497 & -4.28 $\pm$ 0.05 & 115 $\pm$ 48 \\ 
96064 & 54155 & HH Leo & 0.77 & \citep{fuhrmann2004} & 0.496 & -4.36 $\pm$ 0.05 & 213 $\pm$ 81 \\ 
97658 & 54906 & BD+26 2184 & 0.855 & \citep{koen2010} & 0.231 & -4.85 $\pm$ 0.11 &  \nodata \\ 
98649 & 55409 & LTT 4199 & 0.656 & \citep{hog2000} & 0.208 & -4.79 $\pm$ 0.15 &  \nodata \\ 
105631 & 59280 & BD+41 2276 & 0.79 & \citep{kotoneva2006} & 0.307 & -4.63 $\pm$ 0.08 & 1394 $\pm$ 630 \\ 
109272 & 61296 & HR 4779 & 0.831 & \citep{ammler2012} & 0.221 & -4.86 $\pm$ 0.12 &  \nodata \\ 
112733 & 63317 & BD+39 2586 & 0.78 & \citep{hog2000} & 0.398 & -4.48 $\pm$ 0.06 & 534 $\pm$ 219 \\ 
115383 & 64792 & GJ 504 & 0.59 & \citep{chen2000} & 0.287 & -4.50 $\pm$ 0.09 & 618 $\pm$ 390 \\ 
115617 & 64924 & 61 Vir & 0.7 & \citep{oja1993} & 0.198 & -4.86 $\pm$ 0.16 &  \nodata \\ 
120136 & 67275 & $\tau$ Boo & 0.49 & \citep{ducati2002} & 0.136 & -5.01 $\pm$ 0.35 &  \nodata \\ 
120352 & 67412 & BD-00 2743 & 0.75 & \citep{hog2000} & 0.381 & -4.48 $\pm$ 0.06 & 522 $\pm$ 230 \\ 
\hline
\end{tabular}
\label{tbl:chromospheric}
\end{table*}

\begin{table*}
\small
\centering
\contcaption{Chromospheric Activity Index and Ages}
\begin{tabular}{lllccccc}
\hline
{HD} & {HIP} & {Other Name} & {B-V} & {B-V references} & {\HKindex} & {\chrom} & {Age (Myr)} \\
\hline 
128311 & 71395 & HN Boo & 0.995 & \citep{koen2010} & 0.847 & -4.36 $\pm$ 0.03 & 224 $\pm$ 52 \\ 
129333 & 71631 & Ek Dra & 0.59 & \citep{casagrande2011} & 0.685 & -4.02 $\pm$ 0.04 & 10 $\pm$ 4 \\ 
135599 & 74702 & V379 Ser & 0.843 & \citep{koen2010} & 0.655 & -4.29 $\pm$ 0.04 & 129 $\pm$ 40 \\ 
152555 & 82688 & BD-04 4194 & 0.6 & \citep{hog2000} & 0.393 & -4.32 $\pm$ 0.06 & 164 $\pm$ 85 \\ 
189733 & 98505 & V452 Vul & 0.93 & \citep{koen2010} & 0.732 & -4.34 $\pm$ 0.03 & 194 $\pm$ 51 \\ 
202575 & 105038 & LTT 16242 & 1.045 & \citep{koen2010} & 1.034 & -4.34 $\pm$ 0.03 & 195 $\pm$ 41 \\ 
206466 & 107146 & BD+08 3000 & 1.35 & \citep{hog2000} & 0.412 & -5.25 $\pm$ 0.05 &  \nodata \\ 
206860 & 107350 & HN Peg & 0.58 & \citep{hog2000} & 0.324 & -4.42 $\pm$ 0.08 & 340 $\pm$ 201 \\ 
210667 & 109527 & V446 Lac & 0.82 & \citep{fuhrmann2008} & 0.283 & -4.70 $\pm$ 0.09 &  \nodata \\ 
217343 & 113579 & GC 32053 & 0.64 & \citep{hog2000} & 0.556 & -4.17 $\pm$ 0.05 & 44 $\pm$ 18 \\ 
217813 & 113829 & MT Peg & 0.633 & \citep{koen2010} & 0.262 & -4.60 $\pm$ 0.11 & 1150 $\pm$ 730 \\ 
222582 & 116906 & BD-06 6262 & 0.65 & \citep{hog2000} & 0.169 & -4.96 $\pm$ 0.23 &  \nodata \\ 
283750 & 21482 & V833 Tau & 1.12 & \citep{mishenina2008} & 2.983 & -3.99 $\pm$ 0.02 & 8 $\pm$ 2 \\ 
\nodata & 36357 & V376 Gem & 0.96 & \citep{kovtyukh2003} & 0.575 & -4.49 $\pm$ 0.04 & 575 $\pm$ 152 \\ 
\nodata & 115162 & BD+41 4749 & 0.74 & \citep{hog2000} & 0.665 & -4.18 $\pm$ 0.04 & 47 $\pm$ 17 \\ 
\nodata & \nodata & HatP 13 & 0.73 & \citep{hog2000} & 0.173 & -4.99 $\pm$ 0.20 &  \nodata \\ 
\nodata & \nodata & HatP 30 & 0.6 & \citep{hog2000} & 0.233 & -4.65 $\pm$ 0.13 &  \nodata \\ 
\nodata & \nodata & Wasp 12 & 0.57 & \citep{hog2000} & 0.116 & -5.38 $\pm$ 0.71 &  \nodata \\ 
\hline
\end{tabular}
\end{table*}

\begin{table*}
\small
\centering
\caption{Fundamental Atmospheric Parameters for our SEEDS target sample. Note that all K-type stars utilized a shorter wavelength range in the TGVIT solution.}
\begin{tabular}{lllccccccccc}
\hline
{HD} & {HIP} & {Other} & {rv} & {\temp} & {\logg} & {$v_t$ } & {\feh} & {FeI} & {FeII} & {Wavelength} \\
{} & {} & {Name} & {($\frac{km}{sec}$)} & {(K)} & {log$_{10}$($cm/s^2$)} & {($\frac{km}{sec}$)} & {(dex)} & {} & {} & {Region} \\
\hline
166 & 544 & V439 And & -6.3 & 5509 $\pm$ 34 & 4.49 $\pm$ 0.09 & 1.34 $\pm$ 0.15 & 0.00 $\pm$ 0.03 & 107 & 8 & 5500-6968 \\ 
1835 & 1803 & BE Cet & -4.3 & 5765 $\pm$ 55 & 4.52 $\pm$ 0.13 & 1.31 $\pm$ 0.28 & 0.06 $\pm$ 0.06 & 161 & 16 & 4478-6968 \\ 
4128 & 3419 & $\beta$ Cet & 12.4 & 4894 $\pm$ 50 & 2.65 $\pm$ 0.17 & 1.44 $\pm$ 0.22 & -0.08 $\pm$ 0.06 & 125 & 14 & 5500-6968 \\ 
4747 & 3850 & GJ 36 & 9.3 & 5382 $\pm$ 44 & 4.80 $\pm$ 0.10 & 0.89 $\pm$ 0.29 & -0.29 $\pm$ 0.04 & 153 & 14 & 4478-6968 \\ 
4813 & 3909 & 19 Cet & 7.7 & 6170 $\pm$ 34 & 4.28 $\pm$ 0.07 & 1.21 $\pm$ 0.15 & -0.17 $\pm$ 0.03 & 129 & 15 & 4478-6968 \\ 
5608 & 4552 & HR 275 & -25.6 & 4950 $\pm$ 50 & 3.37 $\pm$ 0.16 & 1.24 $\pm$ 0.19 & 0.00 $\pm$ 0.05 & 106 & 7 & 5500-6968 \\ 
7590 & 5944 & V445 And & -13.1 & 5936 $\pm$ 56 & 4.53 $\pm$ 0.12 & 1.29 $\pm$ 0.30 & -0.17 $\pm$ 0.05 & 144 & 13 & 4478-6968 \\ 
7661 & 5938 & EW Cet & 6.9 & 5463 $\pm$ 48 & 4.79 $\pm$ 0.11 & 1.01 $\pm$ 0.27 & -0.02 $\pm$ 0.05 & 153 & 14 & 5500-6968 \\ 
8673 & 6702 & LTT 10515 & 16.4 & 6606 $\pm$ 268 & 4.90 $\pm$ 0.43 & 3.50 $\pm$ 1.15 & 0.11 $\pm$ 0.18 & 99 & 11 & 4478-6968 \\ 
8907 & 6878 & BD+41 283 & 7.8 & 6467 $\pm$ 125 & 4.99 $\pm$ 0.24 & 2.55 $\pm$ 0.54 & -0.05 $\pm$ 0.08 & 128 & 15 & 4478-6968 \\ 
8941 & 6869 & BD+16 154 & 6.5 & 6555 $\pm$ 111 & 4.70 $\pm$ 0.19 & 1.98 $\pm$ 0.38 & 0.22 $\pm$ 0.08 & 133 & 14 & 4478-6968 \\ 
9826 & 7513 & $\upsilon$ And & -30.6 & 6156 $\pm$ 62 & 4.21 $\pm$ 0.13 & 1.53 $\pm$ 0.34 & 0.00 $\pm$ 0.06 & 143 & 14 & 4478-6968 \\ 
10700 & 8102 & $\tau$ Cet & -17.7 & 5347 $\pm$ 31 & 4.81 $\pm$ 0.08 & 0.66 $\pm$ 0.29 & -0.54 $\pm$ 0.30 & 142 & 12 & 4478-6968 \\ 
12039 & 9141 & DK Cet & 5.3 & 5827 $\pm$ 109 & 4.36 $\pm$ 0.26 & 2.53 $\pm$ 0.52 & -0.24 $\pm$ 0.09 & 136 & 13 & 4478-6968 \\ 
13507 & 10321 & V450 And & 5.2 & 5695 $\pm$ 46 & 4.73 $\pm$ 0.10 & 1.13 $\pm$ 0.23 & -0.09 $\pm$ 0.04 & 160 & 13 & 4478-6968 \\ 
14067 & 10657 & HR 665 & -14.0 & 4826 $\pm$ 39 & 2.76 $\pm$ 0.13 & 1.46 $\pm$ 0.23 & -0.25 $\pm$ 0.06 & 147 & 15 & 4478-6968 \\ 
14082B & 10679 & BD+28 382B & 4.1 & 5891 $\pm$ 52 & 4.65 $\pm$ 0.11 & 1.33 $\pm$ 0.30 & -0.07 $\pm$ 0.05 & 148 & 14 & 4478-6968 \\ 
16160 & 12114 & HR 753 & 24.4 & 4858 $\pm$ 69 & 4.89 $\pm$ 0.20 & 0.92 $\pm$ 0.45 & -0.18 $\pm$ 0.06 & 110 & 5 & 5500-6968 \\ 
16760 & 12638 & BD+37 604 & -4.3 & 5596 $\pm$ 31 & 4.69 $\pm$ 0.07 & 0.94 $\pm$ 0.23 & -0.09 $\pm$ 0.04 & 157 & 15 & 4478-6968 \\ 
17925 & 13402 & EP Eri & 17.0 & 5204 $\pm$ 33 & 4.60 $\pm$ 0.09 & 1.37 $\pm$ 0.24 & -0.02 $\pm$ 0.04 & 104 & 6 & 5500-6968 \\ 
18632 & 13976 & BZ Cet & 27.7 & 5020 $\pm$ 66 & 4.88 $\pm$ 0.18 & 1.06 $\pm$ 0.50 & 0.11 $\pm$ 0.07 & 143 & 12 & 5500-6968 \\ 
20630 & 15457 & $\kappa$ Cet & 17.8 & 5750 $\pm$ 38 & 4.69 $\pm$ 0.08 & 1.23 $\pm$ 0.22 & -0.01 $\pm$ 0.04 & 156 & 14 & 4478-6968 \\ 
22781 & 17187 & BD+31 630 & 7.7 & 4981 $\pm$ 44 & 4.84 $\pm$ 0.13 & 0.51 $\pm$ 0.46 & -0.39 $\pm$ 0.03 & 114 & 5 & 5500-6968 \\ 
24916 & 18512 & BD-01 565 & 2.5 & 4725 $\pm$ 164 & 4.81 $\pm$ 0.52 & 1.33 $\pm$ 0.53 & -0.11 $\pm$ 0.14 & 112 & 4 & 5500-6968 \\ 
25457 & 18859 & BD-00 423 & 17.1 & 6562 $\pm$ 197 & 4.59 $\pm$ 0.34 & 2.64 $\pm$ 0.54 & 0.06 $\pm$ 0.12 & 125 & 12 & 4478-6968 \\ 
25665 & 19422 & BD+69 238 & -15.1 & 5022 $\pm$ 36 & 4.79 $\pm$ 0.10 & 0.97 $\pm$ 0.28 & -0.07 $\pm$ 0.04 & 105 & 6 & 5500-6968 \\ 
25998 & 19335 & V582 Per & 25.5 & 6204 $\pm$ 137 & 4.34 $\pm$ 0.25 & 2.19 $\pm$ 0.52 & -0.03 $\pm$ 0.10 & 130 & 10 & 4478-6968 \\ 
29697 & 21818 & V834 Tau & 1.6 & 4516 $\pm$ 152 & 3.93 $\pm$ 0.53 & 2.28 $\pm$ 0.56 & -0.53 $\pm$ 0.15 & 98 & 6 & 5500-6968 \\ 
31000 & 22776 & V536 Aur & -5.1 & 5363 $\pm$ 44 & 4.67 $\pm$ 0.11 & 1.24 $\pm$ 0.26 & -0.09 $\pm$ 0.05 & 148 & 14 & 4478-6968 \\ 
36869 & \nodata & AH Lep & 23.9 & 6533 $\pm$ 297 & 6.26 $\pm$ 0.64 & 5.42 $\pm$ 4.81 & 0.26 $\pm$ 0.21 & \nodata & \nodata & 4478-6968 \\ 
37394 & 26779 & V538 Aur & -0.4 & 5309 $\pm$ 26 & 4.65 $\pm$ 0.07 & 1.08 $\pm$ 0.14 & 0.06 $\pm$ 0.03 & 108 & 7 & 5500-6968 \\ 
38393 & 27072 & LTT 2364 & -11.1 & 6241 $\pm$ 54 & 4.19 $\pm$ 0.10 & 1.46 $\pm$ 0.21 & -0.16 $\pm$ 0.04 & 120 & 15 & 4478-6968 \\ 
39587 & 27913 & chi01 Ori & -13.5 & 5981 $\pm$ 42 & 4.62 $\pm$ 0.09 & 1.37 $\pm$ 0.22 & -0.09 $\pm$ 0.04 & 145 & 15 & 4478-6968 \\ 
40774 & 28526 & BD+09 1055 & -14.5 & 5088 $\pm$ 42 & 2.92 $\pm$ 0.14 & 1.42 $\pm$ 0.20 & -0.34 $\pm$ 0.06 & 159 & 16 & 4478-6968 \\ 
41593 & 28954 & V1386 Ori & -10.8 & 5321 $\pm$ 29 & 4.72 $\pm$ 0.07 & 1.25 $\pm$ 0.17 & -0.03 $\pm$ 0.03 & 103 & 7 & 5500-6968 \\ 
43162 & 29568 & GJ 3389 & 19.9 & 5607 $\pm$ 36 & 4.72 $\pm$ 0.09 & 1.31 $\pm$ 0.23 & -0.11 $\pm$ 0.04 & 152 & 15 & 4478-6968 \\ 
59747 & 36704 & DX Lyn & -17.4 & 5120 $\pm$ 43 & 4.84 $\pm$ 0.11 & 1.14 $\pm$ 0.29 & -0.06 $\pm$ 0.04 & 128 & 11 & 4478-6968 \\ 
60737 & 37170 & GC 10209 & 6.9 & 5832 $\pm$ 63 & 4.54 $\pm$ 0.13 & 1.62 $\pm$ 0.41 & -0.25 $\pm$ 0.06 & 134 & 14 & 4478-6968 \\ 
61606 & 37349 & V869 Mon & -19.4 & 5034 $\pm$ 44 & 4.96 $\pm$ 0.06 & 1.15 $\pm$ 0.04 & -0.02 $\pm$ 0.04 & 101 & 6 & 5500-6968 \\ 
63433 & 38228 & V377 Gem & -14.5 & 5660 $\pm$ 50 & 4.76 $\pm$ 0.12 & 1.21 $\pm$ 0.28 & -0.08 $\pm$ 0.05 & 155 & 14 & 4478-6968 \\ 
68988 & 40687 & BD+61 1038 & -69.3 & 5864 $\pm$ 41 & 4.40 $\pm$ 0.10 & 1.13 $\pm$ 0.20 & 0.19 $\pm$ 0.05 & 157 & 16 & 4478-6968 \\ 
69830 & 40693 & LHS 245 & 28.0 & 5442 $\pm$ 41 & 4.79 $\pm$ 0.09 & 0.83 $\pm$ 0.29 & -0.08 $\pm$ 0.05 & 161 & 14 & 4478-6968 \\ 
72760 & 42074 & BD-00 2024 & 34.3 & 5295 $\pm$ 44 & 4.78 $\pm$ 0.11 & 1.02 $\pm$ 0.30 & -0.04 $\pm$ 0.05 & 157 & 14 & 4478-6968 \\ 
72905 & 42438 & 3 UMa & -13.2 & 5866 $\pm$ 42 & 4.66 $\pm$ 0.09 & 1.37 $\pm$ 0.29 & -0.15 $\pm$ 0.04 & 145 & 13 & 4478-6968 \\ 
73350 & 42333 & V401 Hya & 34.5 & 5778 $\pm$ 39 & 4.59 $\pm$ 0.08 & 1.13 $\pm$ 0.25 & -0.02 $\pm$ 0.05 & 156 & 14 & 4478-6968 \\ 
75732 & 43587 & LHS 2062 & 26.6 & 5316 $\pm$ 79 & 4.84 $\pm$ 0.20 & 0.82 $\pm$ 0.48 & 0.26 $\pm$ 0.08 & 153 & 14 & 4478-6968 \\ 
76151 & 43726 & BD-04 2490 & 30.9 & 5794 $\pm$ 36 & 4.69 $\pm$ 0.08 & 1.02 $\pm$ 0.22 & 0.02 $\pm$ 0.04 & 157 & 15 & 4478-6968 \\ 
79555 & 45383 & HEI 350 & 11.3 & 4871 $\pm$ 69 & 4.68 $\pm$ 0.21 & 1.28 $\pm$ 0.39 & -0.18 $\pm$ 0.06 & 103 & 6 & 5500-6968 \\ 
81040 & 46076 & BD+20 2314 & 48.9 & 5747 $\pm$ 32 & 4.71 $\pm$ 0.07 & 0.91 $\pm$ 0.21 & -0.14 $\pm$ 0.03 & 153 & 14 & 4478-6968 \\ 
82106 & 46580 & LHS2147 & 28.0 & 4836 $\pm$ 74 & 4.77 $\pm$ 0.22 & 1.20 $\pm$ 0.46 & -0.11 $\pm$ 0.07 & 106 & 5 & 5500-6968 \\ 
82443 & 46843 & DX Leo & 13.0 & 5361 $\pm$ 41 & 4.50 $\pm$ 0.11 & 1.52 $\pm$ 0.19 & -0.14 $\pm$ 0.04 & 100 & 7 & 5500-6968 \\ 
87424 & 49366 & V417 Hya & -12.4 & 5106 $\pm$ 53 & 5.02 $\pm$ 0.14 & 0.76 $\pm$ 0.05 & -0.16 $\pm$ 0.05 & 154 & 10 & 4478-6968 \\ 
89744 & 50786 & BD+41 2076 & -7.7 & 6139 $\pm$ 71 & 3.95 $\pm$ 0.15 & 1.82 $\pm$ 0.35 & 0.00 $\pm$ 0.06 & 141 & 15 & 4478-6968 \\ 
90839 & 51459 & GJ 395 & 7.4 & 6148 $\pm$ 32 & 4.39 $\pm$ 0.07 & 1.24 $\pm$ 0.16 & -0.13 $\pm$ 0.25 & 140 & 16 & 4478-6968 \\ 
94765 & 53486 & GY Leo & 4.8 & 5071 $\pm$ 48 & 4.69 $\pm$ 0.13 & 1.17 $\pm$ 0.25 & -0.04 $\pm$ 0.04 & 106 & 6 & 5500-6968 \\ 
95174 & \nodata & LTT 12936 & -3.6 & 4748 $\pm$ 74 & 4.68 $\pm$ 0.23 & 1.20 $\pm$ 0.45 & -0.13 $\pm$ 0.07 & 106 & 6 & 5500-6968 \\ 
96064 & 54155 & HH Leo & 17.6 & 5448 $\pm$ 51 & 4.81 $\pm$ 0.13 & 1.24 $\pm$ 0.29 & -0.06 $\pm$ 0.05 & 149 & 13 & 4478-6968 \\ 
96167 & 54195 & BD-09 3201 & 10.9 & 5748 $\pm$ 44 & 4.17 $\pm$ 0.10 & 1.20 $\pm$ 0.20 & 0.22 $\pm$ 0.05 & 157 & 16 & 4478-6968 \\ 
97658 & 54906 & BD+26 2184 & -2.6 & 5174 $\pm$ 37 & 4.65 $\pm$ 0.09 & 1.06 $\pm$ 0.23 & -0.38 $\pm$ 0.03 & 109 & 7 & 5500-6968 \\ 
98649 & 55409 & LTT 4199 & 3.3 & 5683 $\pm$ 33 & 4.55 $\pm$ 0.07 & 0.98 $\pm$ 0.20 & -0.13 $\pm$ 0.03 & 159 & 15 & 4478-6968 \\ 
102956 & 57820 & BD+58 1340 & -25.6 & 5065 $\pm$ 50 & 3.51 $\pm$ 0.18 & 1.28 $\pm$ 0.19 & 0.02 $\pm$ 0.06 & 105 & 8 & 5500-6968 \\ 
109272 & 61296 & HR 4779 & -15.4 & 5059 $\pm$ 33 & 3.41 $\pm$ 0.09 & 1.11 $\pm$ 0.15 & -0.34 $\pm$ 0.04 & 147 & 15 & 4478-6968 \\ 
\hline
\end{tabular}
\label{tbl:atmo_results}
\end{table*}

\begin{table*}
\small
\centering
\caption{Fundamental Atmospheric Parameters for our SEEDS target sample. Note that all K-type stars utilized a shorter wavelength range in the TGVIT solution.}
\begin{tabular}{lllccccccccc}
\hline
{HD} & {HIP} & {Other} & {rv} & {\temp} & {\logg} & {$v_t$ } & {\feh} & {FeI} & {FeII} & {Wavelength} \\
{} & {} & {Name} & {($\frac{km}{sec}$)} & {(K)} & {log$_{10}$($cm/s^2$)} & {($\frac{km}{sec}$)} & {(dex)} & {} & {} & {Region} \\
\hline
112733 & 63317 & BD+39 2586 & -3.6 & 5401 $\pm$ 38 & 4.76 $\pm$ 0.10 & 1.21 $\pm$ 0.27 & -0.18 $\pm$ 0.04 & 149 & 12 & 4478-6968 \\ 
115383 & 64792 & GJ 504 & -27.6 & 6063 $\pm$ 62 & 4.38 $\pm$ 0.13 & 1.30 $\pm$ 0.24 & 0.07 $\pm$ 0.05 & 146 & 14 & 4478-6968 \\ 
115617 & 64924 & 61 Vir & -8.8 & 5550 $\pm$ 32 & 4.46 $\pm$ 0.09 & 0.91 $\pm$ 0.18 & -0.09 $\pm$ 0.04 & 157 & 14 & 4478-6968 \\ 
120136 & 67275 & $\tau$ Boo & -17.4 & 6584 $\pm$ 132 & 4.77 $\pm$ 0.23 & 2.01 $\pm$ 0.49 & 0.23 $\pm$ 0.09 & 133 & 13 & 4478-6968 \\ 
120352 & 67412 & BD-00 2743 & -14.0 & 5610 $\pm$ 23 & 4.62 $\pm$ 0.06 & 1.21 $\pm$ 0.16 & -0.01 $\pm$ 0.03 & 112 & 7 & 5500-6968 \\ 
128167 & 71284 & $\sigma$ Boo & -0.8 & 6547 $\pm$ 105 & 3.95 $\pm$ 0.16 & 1.86 $\pm$ 0.37 & -0.54 $\pm$ 0.06 & 88 & 15 & 4478-6968 \\ 
128311 & 71395 & HN Boo & -11.6 & 5025 $\pm$ 44 & 4.97 $\pm$ 0.07 & 1.20 $\pm$ 0.05 & 0.06 $\pm$ 0.06 & 104 & 6 & 5500-6968 \\ 
129333 & 71631 & Ek Dra & -21.3 & 5853 $\pm$ 113 & 4.60 $\pm$ 0.23 & 2.51 $\pm$ 0.52 & -0.17 $\pm$ 0.09 & 133 & 14 & 4478-6968 \\ 
135599 & 74702 & V379 Ser & -3.1 & 5229 $\pm$ 21 & 4.58 $\pm$ 0.05 & 1.14 $\pm$ 0.16 & -0.15 $\pm$ 0.02 & 107 & 7 & 5500-6968 \\ 
145229 & 79165 & BD+11 2925 & -38.1 & 5950 $\pm$ 40 & 4.60 $\pm$ 0.09 & 1.21 $\pm$ 0.24 & -0.24 $\pm$ 0.04 & 139 & 12 & 4478-6968 \\ 
152555 & 82688 & BD-04 4194 & -16.3 & 6201 $\pm$ 107 & 4.80 $\pm$ 0.21 & 2.12 $\pm$ 0.51 & -0.02 $\pm$ 0.08 & 135 & 13 & 4478-6968 \\ 
189733 & 98505 & V452 Vul & 1.4 & 5092 $\pm$ 48 & 4.79 $\pm$ 0.16 & 1.11 $\pm$ 0.27 & -0.07 $\pm$ 0.04 & 104 & 5 & 5500-6968 \\ 
199665 & 103527 & 18 Del & 7.6 & 4965 $\pm$ 40 & 3.06 $\pm$ 0.13 & 1.34 $\pm$ 0.22 & -0.08 $\pm$ 0.06 & 144 & 16 & 4478-6968 \\ 
202575 & 105038 & LTT 16242 & -18.4 & 4790 $\pm$ 63 & 4.78 $\pm$ 0.18 & 1.29 $\pm$ 0.48 & -0.27 $\pm$ 0.06 & 105 & 6 & 5500-6968 \\ 
206466 & 107146 & BD+08 3000 & 2.1 & 5887 $\pm$ 41 & 4.44 $\pm$ 0.10 & 1.32 $\pm$ 0.17 & -0.10 $\pm$ 0.04 & 105 & 7 & 5500-6968 \\ 
206860 & 107350 & HN Peg & -17.4 & 6001 $\pm$ 54 & 4.61 $\pm$ 0.11 & 1.43 $\pm$ 0.39 & -0.14 $\pm$ 0.06 & 140 & 13 & 4478-6968 \\ 
210667 & 109527 & V446 Lac & -21.2 & 5338 $\pm$ 41 & 4.61 $\pm$ 0.10 & 1.06 $\pm$ 0.26 & 0.03 $\pm$ 0.05 & 154 & 14 & 4478-6968 \\ 
212698 & \nodata & 53 Aqr A & -1.0 & 5888 $\pm$ 51 & 4.59 $\pm$ 0.10 & 1.40 $\pm$ 0.32 & -0.17 $\pm$ 0.05 & 142 & 13 & 4478-6968 \\ 
217343 & 113579 & GC 32053 & 5.6 & 5736 $\pm$ 79 & 4.77 $\pm$ 0.17 & 1.60 $\pm$ 0.44 & -0.13 $\pm$ 0.07 & 145 & 14 & 4478-6968 \\ 
217813 & 113829 & MT Peg & 0.1 & 5874 $\pm$ 31 & 4.57 $\pm$ 0.07 & 1.23 $\pm$ 0.19 & -0.03 $\pm$ 0.03 & 159 & 16 & 4478-6968 \\ 
220182 & 115331 & V453 And & 1.3 & 5287 $\pm$ 45 & 4.67 $\pm$ 0.11 & 1.23 $\pm$ 0.31 & -0.13 $\pm$ 0.05 & 150 & 14 & 4478-6968 \\ 
222582 & 116906 & BD-06 6262 & 9.8 & 5733 $\pm$ 29 & 4.42 $\pm$ 0.07 & 1.06 $\pm$ 0.16 & -0.10 $\pm$ 0.03 & 157 & 16 & 4478-6968 \\ 
283750 & 21482 & V833 Tau & 35.8 & 4787 $\pm$ 102 & 4.34 $\pm$ 0.35 & 2.49 $\pm$ 0.52 & -0.11 $\pm$ 0.09 & 94 & 5 & 5500-6968 \\ 
\nodata & 36357 & V376 Gem & -4.6 & 4992 $\pm$ 50 & 4.72 $\pm$ 0.14 & 1.22 $\pm$ 0.33 & -0.22 $\pm$ 0.04 & 106 & 6 & 5500-6968 \\ 
\nodata & 97657 & HatP 11 & -60.3 & 4795 $\pm$ 61 & 4.87 $\pm$ 0.19 & 1.11 $\pm$ 0.49 & 0.15 $\pm$ 0.07 & 98 & 6 & 5500-6968 \\ 
\nodata & 115162 & BD+41 4749 & -20.7 & 5508 $\pm$ 29 & 4.76 $\pm$ 0.07 & 1.19 $\pm$ 0.27 & -0.09 $\pm$ 0.04 & 156 & 15 & 4478-6968 \\ 
\nodata & \nodata & HatP 13 & 12.5 & 5722 $\pm$ 61 & 4.33 $\pm$ 0.15 & 1.15 $\pm$ 0.27 & 0.33 $\pm$ 0.07 & 157 & 16 & 4478-6968 \\ 
\nodata & \nodata & HatP 17 & 18.8 & 5235 $\pm$ 44 & 4.69 $\pm$ 0.11 & 0.75 $\pm$ 0.41 & -0.09 $\pm$ 0.06 & 160 & 14 & 4478-6968 \\ 
\nodata & \nodata & HatP 30 & 43.7 & 6228 $\pm$ 55 & 4.24 $\pm$ 0.10 & 1.17 $\pm$ 0.28 & -0.04 $\pm$ 0.05 & 140 & 16 & 4478-6968 \\ 
\nodata & \nodata & HatP 6 & -22.9 & 6629 $\pm$ 105 & 4.20 $\pm$ 0.15 & 2.56 $\pm$ 0.53 & -0.40 $\pm$ 0.06 & 93 & 15 & 4478-6968 \\ 
\nodata & \nodata & Wasp 12 & 18.4 & 6217 $\pm$ 45 & 4.18 $\pm$ 0.08 & 1.51 $\pm$ 0.23 & 0.01 $\pm$ 0.04 & 143 & 16 & 4478-6968 \\ 
\hline
\end{tabular}
\end{table*}

\end{document}